\documentclass[aps,prl,superscriptaddress]{revtex4}
\usepackage{epsfig}
\usepackage{german}
\usepackage{curves}
\curvewarnfalse

\begin{document}

\title{Neutron Radiography Analysis of a Transient Liquid Phase Joint}

\author{H. Ballhausen}
\affiliation{Physical Institute, University of Heidelberg, Philosophenweg 12, 69120 Heidelberg, Germany}
\email{ballhausen@physi.uni-heidelberg.de}

\author{H. Abele}
\affiliation{Physical Institute, University of Heidelberg, Philosophenweg 12, 69120 Heidelberg, Germany}

\author{R. S. Eccleston}
\affiliation{Materials and Engineering Research Institute, Sheffield Hallam University, Sheffield, UK}

\author{R. G"ahler}
\affiliation{Institut Laue Langevin, 6 rue Jules Horowitz BP 156, 38042 Grenoble Cedex 9, France}

\author{A. J. Smith}
\affiliation{Materials and Engineering Research Institute, Sheffield Hallam University, Sheffield, UK}

\author{A. Steuwer}
\affiliation{FaME38 at the ESRF-ILL, 6 rue Jules Horowitz, 38042 Grenoble Cedex 9, France}

\author{A. Van Overberghe}
\affiliation{Physical Institute, University of Heidelberg, Philosophenweg 12, 69120 Heidelberg, Germany}
\affiliation{Institut Laue Langevin, 6 rue Jules Horowitz BP 156, 38042 Grenoble Cedex 9, France}

\begin{abstract}
\noindent
Neutron radiography in many cases is the only non-destructive technique available for the analysis of a wide range
of samples from metallurgy, materials engineering and materials testing. In this paper the potential of the
technique is illustrated for a transient liquid phase (TLP) joint.
\\
TLP bonding produces interface free and stress free joints.
The quality and properties of the joint depend on the diffusion of an interlayer into the base material.
A TLP joint is visualised and the diffusion profile of the boron contained in the bonding additives is determined.
Parameters of the bonding process are determined quantitatively from this profile, and flaws in the joint are detected.

\end{abstract}


\maketitle

\section{I. Introduction}

\noindent
Neutron radiography \cite{NR1,NR3,NR4,neutrograph} is a non-destructive technique and
produces attenuation images similar to X-ray methods. Neutrons have the advantage over X-rays that they easily
penetrate metals while showing characterisic cross-sections for different elements.
With this complementary and unique probe neutron radiography is able to resolve
sub-$10^{-3}$ variations of contrast on a sub-millimeter scale.
In this paper we report the first demonstration of a non-destructive quantitative method for the analysis
of a TLP joint.

\bigskip
\noindent
Transient liquid phase bonding \cite{tlp} is an advanced bonding technique.
In most simple terms, the bonding process works as follows: A thin foil is
placed between the surfaces to be bonded. It contains melting point depressants
which lower the melting point of the base material. At constant temperature
those parts of the assembly are liquid where the concentration of the additive
is above a certain threshold. As the additive diffuses into the base material
the liquid phase extends, hence the name of the method. When through diffusion
the overall concentration of the additive falls below the threshold the material
undergoes isothermal solidification.

\bigskip
\twocolumngrid

\noindent
TLP joints (see fig. 1)
are interface free and virtually stress free due to their isothermal creation.
The more even the diffusion of the additive is, and the lower its final concentration,
the closer are the properties of the joint to those of the base material. Only then has the
joint favourable mechanical properties, whereas regions of higher concentration of the
additive are prone to embrittlement through secondary phases. 
It would therefore be highly advantageous to be able to quantitatively measure the
spatial distribution of the additives around the joint.

\bigskip
\noindent
\begin{center}
\epsfig{file=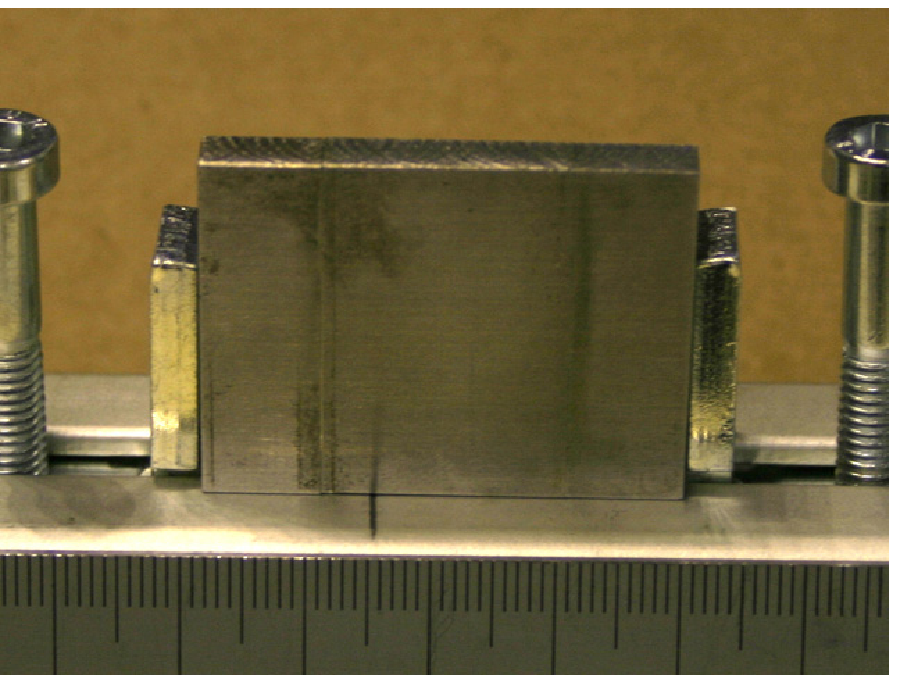,width=5.5cm}\\
Fig. 1: sample featuring a TLP joint
\end{center}

\onecolumngrid

\noindent
Traditional metallographic (destructive) characterisation techniques
cannot reveal this distribution for an 

\twocolumngrid

\noindent
individual specimen \cite{altemethode}.
Upon interruption of the bonding process, the sample is cut to investigate the cross-section using optical methods. The phases can be distinguished by a
microscope, but it is very difficult to obtain a quantitative, rather than
a qualitative diffusion profile. This time consuming method has to be repeated
several times to get a time resolved analysis of the bonding process.\\
In contrast neutron radiography delivers a visualisation of the diffusion
profile (see fig. 2) within milliseconds at any stage of the bonding process
without altering the sample or interfering with the process.

\bigskip
\noindent
\begin{center}
\epsfig{file=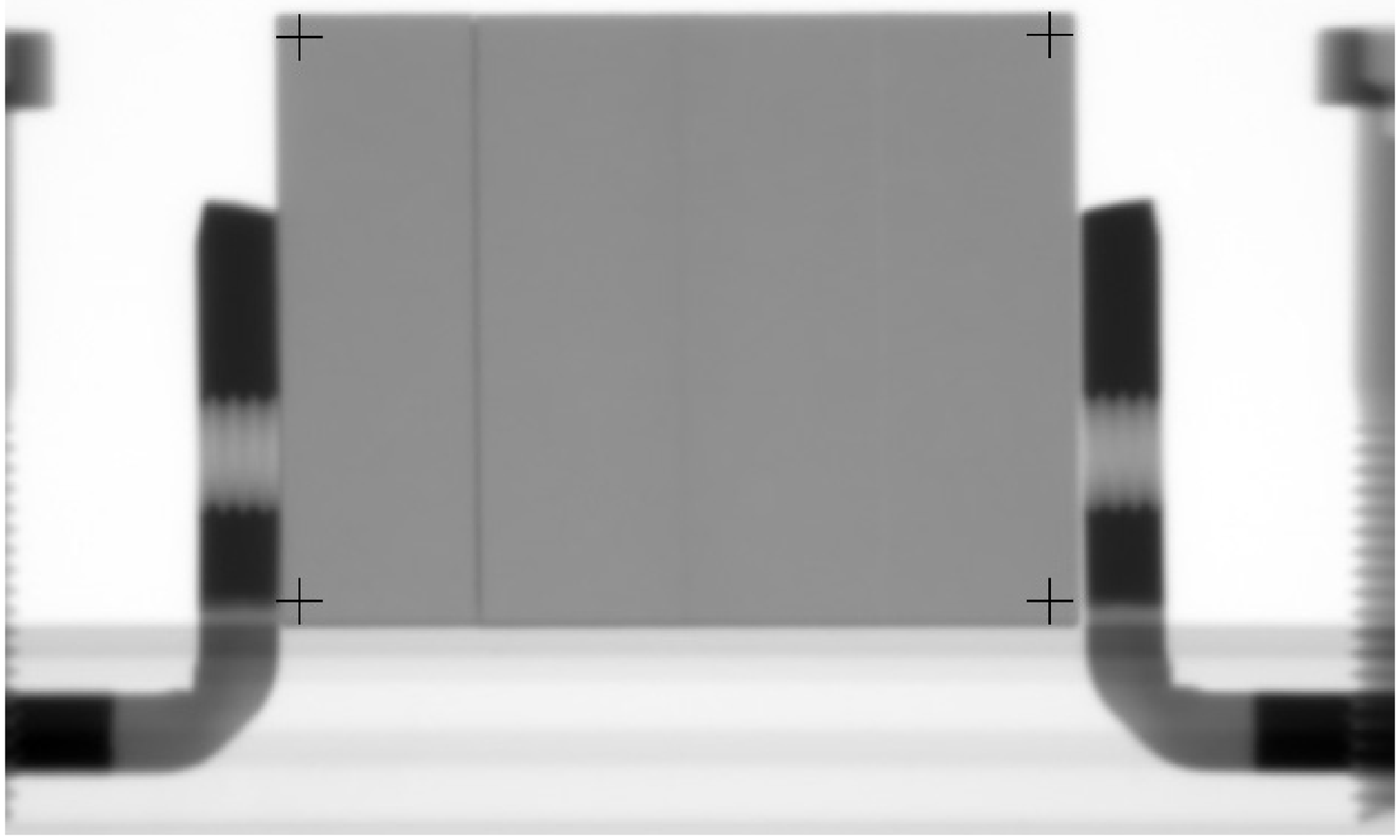,width=5.5cm}\\
Fig. 2: neutron radiography image\\
of the sample in fig. 1
\end{center}

\newpage
\onecolumngrid

\section{II. Transient Liquid Phase Bonding}

\noindent
Consider the binary phase diagram shown in figure 3. It describes the
phase states of a two component alloy. This may vary in composition
from pure base material (B, right) to pure additives (A, left).

\bigskip
\twocolumngrid

\noindent
In this case the additives lower the melting point of the base material.
Temperature is shown on the vertical axis. As one can see the composition
can be solid and liquid at the same temperature depending on the
concentration of the additive within the base material.

\bigskip
\noindent
Consider two isothermal processes
which transform the alloy from one phase into another: 

\bigskip
\noindent
1. Assume that a solid piece of A is embodied in B. At its surface there
is a steep concentration gradient. Solid state diffusion of A into B will
create a gradient between the liquidi $C_{\alpha L}$ and $C_{L \alpha}$.
The liquid phase around B will further dissolute the remaining solid B.
This process is the isothermal solution of A (left arrow).

\bigskip
\noindent
2. Another
process is the isothermal solidification of B (right arrow). Here B diffuses
into an A-rich liquid until its concentration falls below the critical
liquidus $C_{\beta L}$. 

~

\bigskip
\noindent
TLP bonding consists of a combination of three such equithermal processes:
liquidification of A and B and solidification of B. A thin layer of A
(typically $20 \mu m \, ... \, 50 \mu m$) is placed between the two pieces of B
to be bonded. The assembly is heated to typically $800^\circ C ... 1300^\circ C$.
Due to surface interdiffusion there are two thin liquid layers between the
four surfaces between the three bodies (figure 4a). The following
bonding process is ideally divided into three stages:

\bigskip
\noindent
Thermodynamic forces will act to level the steep gradient of concentration
within the liquid layers. At their border additional material is dissolved
and enters the liquid phase to keep the concentrations $C_{L \alpha}$
and $C_{L \beta}$ respectively at the border of the liquid constant.
This stage is called dissolution of the interlayer and is typically 
completed within some seconds (figure 4b).

\bigskip
\noindent
After the interlayer has been completely dissolved the liquid phase 
continues to extend outwards. The concentration may now fall below 
$C_{L \alpha}$ everywhere in the liquid phase and the concentration               
gradient will eventually level and settle down at $C_{L \beta}$.
The lower gradient slows down this second stage considerably which  
typically takes some minutes (figure 4c).

\bigskip
\begin{center}
\unitlength0.5cm
\begin{picture}(16,16)
\thicklines

\put(1,1.5){\vector(0,1){13}}
\put(1,1.5){\line(1,0){14}}
\put(15,1.5){\line(0,1){13}}

\put(0.8,14.7){$T$}
\put(0.8,0.5){A}
\put(14.8,0.5){B}

\thinlines

\put(5,7){\line(1,0){7}}
\qbezier(1.5,1.5)(2.5,7)(5,7)
\qbezier(14.5,1.5)(14.5,7)(12,7)
\put(1,11){\line(1,-1){4}}
\put(1,11){\line(2,-1){8}}
\put(9,7){\line(1,1){6}}
\put(12,7){\line(1,2){3}}

\put(6,12){$liquid~phase~(L)$}
\put(6,3){$solid~phase~(\alpha+\beta)$}

\thicklines
\put(2,9){\vector(1,0){5.2}}
\put(9.2,9){\vector(1,0){4.8}}

\put(5.1,10.1){\tiny{isothermal}}
\put(5.1,9.7){\tiny{liquidifica-}}
\put(5.1,9.3){\tiny{tion of A}}

\put(9.2,10.1){\tiny{isothermal}}
\put(9.2,9.7){\tiny{solidifica-}}
\put(9.2,9.3){\tiny{tion of B}}

\thinlines

\put(1.3,11.2){$Tm_A$}
\put(13.3,13.2){$Tm_B$}

\put(2.5,7.3){$\alpha$}
\put(5.5,7.3){$\alpha + L$}
\put(10.1,7.3){$L + \beta$}
\put(13.5,7.3){$\beta$}

\put(3,1.3){\line(0,1){0.2}}
\put(5,1.3){\line(0,1){0.2}}
\put(11,1.3){\line(0,1){0.2}}
\put(13,1.3){\line(0,1){0.2}}
\put(2.5,0.5){$C_{\alpha L}$}
\put(4.5,0.5){$C_{L \alpha}$}
\put(10.5,0.5){$C_{L \beta}$}
\put(12.5,0.5){$C_{\beta L}$}

\put(2.85,8.82){$\circ$}
\put(4.85,8.82){$\circ$}
\put(10.85,8.82){$\circ$}
\put(12.85,8.82){$\circ$}

\end{picture}\\
Fig. 3: binary phase diagram\\
(~horizontal axis: percentage of (A)dditive vs.\\
(B)ase material, vertical axis: temperature~)
\end{center}

\bigskip
\begin{center}
\unitlength0.5cm
\begin{picture}(16,20)

\thicklines
\put(1,15){\line(1,0){14}}
\put(1,17){\line(1,0){14}}
\put(1,15){\line(0,1){2}}
\put(15,15){\line(0,1){2}}

\put(7.5,16.7){$\cdot\!\!\cdot\!\!\cdot\!\!\cdot\!\!\cdot\!\!\cdot \cdot$}
\put(7.5,16.6){$\cdot\!\!\cdot\!\!\cdot\!\!\cdot\!\!\cdot\!\!\cdot\!\!\cdot$}
\put(7.38,16.5){$\cdot \!\cdot\!\!\cdot\!\!\cdot\!\!\cdot\!\!\cdot\!\!\cdot$}
\put(7.5,16.4){$\cdot\!\!\cdot\!\!\cdot\!\!\cdot\!\!\cdot\!\!\cdot\!\!\cdot$}
\put(7.5,16.3){$\cdot\!\!\cdot\!\!\cdot\!\!\cdot\!\!\cdot\!\!\cdot\!\!\cdot$}
\put(7.5,16.2){$\cdot\!\!\cdot\!\!\cdot\!\!\cdot\!\!\cdot\!\!\cdot\!\!\cdot$}
\put(7.5,16.1){$\cdot\!\!\cdot\!\!\cdot\!\!\cdot\!\!\cdot\!\!\cdot\! \cdot$}
\put(7.38,16.0){$\cdot \!\cdot\!\!\cdot\!\!\cdot\!\!\cdot\!\!\cdot\!\!\cdot$}
\put(7.5,15.9){$\cdot\!\!\cdot\!\!\cdot\!\!\cdot\!\!\cdot\!\!\cdot\!\!\cdot$}
\put(7.5,15.8){$\cdot\!\!\cdot\!\!\cdot\!\!\cdot\!\!\cdot\!\!\cdot\!\!\cdot$}
\put(7.5,15.7){$\cdot\!\!\cdot\!\!\cdot\!\!\cdot\!\!\cdot\!\!\cdot\!\!\cdot$}
\put(7.5,15.6){$\cdot\!\!\cdot\!\!\cdot\!\!\cdot\!\!\cdot\!\!\cdot\!\!\cdot$}
\put(7.5,15.5){$\cdot\!\!\cdot\!\!\cdot\!\!\cdot\!\!\cdot\!\!\cdot \cdot$}
\put(7.26,15.4){$\cdot \cdot\!\!\cdot\!\!\cdot\!\!\cdot\!\!\cdot\!\!\cdot$}
\put(7.5,15.3){$\cdot\!\!\cdot\!\!\cdot\!\!\cdot\!\!\cdot\!\!\cdot\!\!\cdot$}
\put(7.5,15.2){$\cdot\!\!\cdot\!\!\cdot\!\!\cdot\!\!\cdot\!\!\cdot\!\!\cdot$}
\put(7.5,15.1){$\cdot\!\!\cdot\!\!\cdot\!\!\cdot\!\!\cdot\!\!\cdot\! \cdot$}
\put(7.5,15.0){$\cdot\!\!\cdot\!\!\cdot\!\!\cdot\!\!\cdot\!\!\cdot\!\!\cdot$}
\put(7.5,14.9){$\cdot\!\!\cdot\!\!\cdot\!\!\cdot\!\!\cdot\!\!\cdot\!\!\cdot$}

\thinlines
\put(4,17.2){B}
\put(7.7,17.2){A}
\put(12,17.2){B}
\put(5,18.5){\tiny{liquid interlayers at surfaces}}
\put(9.0,18.35){\vector(-1,-2){0.6}}
\put(6.8,18.35){\vector(1,-2){0.6}}
\put(1,14){a) initial state}

\thicklines
\put(1,9){\line(1,0){14}}
\put(1,11){\line(1,0){14}}
\put(1,9){\line(0,1){2}}
\put(15,9){\line(0,1){2}}

\put(07.71,09.72){$\cdot$} \put(07.96,09.86){$\cdot$} \put(08.19,09.09){$\cdot$} \put(08.25,08.92){$\cdot$} \put(09.07,08.92){$\cdot$} \put(09.03,09.09){$\cdot$} \put(07.53,09.58){$\cdot$}
\put(10.28,09.99){$\cdot$} \put(06.77,09.74){$\cdot$} \put(10.57,10.56){$\cdot$} \put(08.61,09.54){$\cdot$} \put(06.43,10.53){$\cdot$} \put(07.73,09.16){$\cdot$} \put(08.30,10.66){$\cdot$}
\put(07.78,09.52){$\cdot$} \put(07.99,09.10){$\cdot$} \put(09.57,10.05){$\cdot$} \put(08.45,09.89){$\cdot$} \put(08.20,08.92){$\cdot$} \put(06.03,10.00){$\cdot$} \put(07.97,10.21){$\cdot$}
\put(07.46,10.24){$\cdot$} \put(06.96,10.01){$\cdot$} \put(04.78,09.33){$\cdot$} \put(09.96,09.89){$\cdot$} \put(12.22,09.95){$\cdot$} \put(08.24,09.68){$\cdot$} \put(06.50,09.80){$\cdot$}
\put(07.59,10.67){$\cdot$} \put(08.72,08.94){$\cdot$} \put(08.80,09.09){$\cdot$} \put(07.63,10.67){$\cdot$} \put(08.49,09.69){$\cdot$} \put(08.00,10.59){$\cdot$} \put(06.54,09.59){$\cdot$}
\put(09.22,10.56){$\cdot$} \put(07.76,09.90){$\cdot$} \put(07.51,09.44){$\cdot$} \put(07.53,09.84){$\cdot$} \put(07.08,09.00){$\cdot$} \put(06.36,10.35){$\cdot$} \put(07.87,10.03){$\cdot$}
\put(05.67,10.09){$\cdot$} \put(07.46,09.77){$\cdot$} \put(08.35,10.18){$\cdot$} \put(06.48,09.37){$\cdot$} \put(06.53,09.38){$\cdot$} \put(05.51,10.56){$\cdot$} \put(07.96,09.15){$\cdot$}
\put(08.35,09.51){$\cdot$} \put(07.77,10.33){$\cdot$} \put(08.14,09.08){$\cdot$} \put(06.94,10.69){$\cdot$} \put(07.45,09.31){$\cdot$} \put(09.21,09.21){$\cdot$} \put(06.48,10.16){$\cdot$}
\put(09.82,10.60){$\cdot$} \put(09.25,09.17){$\cdot$} \put(06.22,10.31){$\cdot$} \put(06.04,09.46){$\cdot$} \put(06.42,10.03){$\cdot$} \put(08.01,10.05){$\cdot$} \put(09.61,09.18){$\cdot$}
\put(08.74,09.78){$\cdot$} \put(11.34,10.10){$\cdot$} \put(07.45,09.08){$\cdot$} \put(07.15,10.64){$\cdot$} \put(08.18,09.24){$\cdot$} \put(06.16,09.23){$\cdot$} \put(05.80,09.50){$\cdot$}
\put(09.83,10.56){$\cdot$} \put(11.13,09.85){$\cdot$} \put(04.06,10.14){$\cdot$} \put(09.21,09.65){$\cdot$} \put(11.26,09.88){$\cdot$} \put(08.20,10.14){$\cdot$} \put(08.12,09.93){$\cdot$}
\put(08.01,09.74){$\cdot$} \put(08.48,10.18){$\cdot$} \put(07.89,09.76){$\cdot$} \put(09.26,10.14){$\cdot$} \put(08.12,09.43){$\cdot$} \put(03.53,09.97){$\cdot$} \put(10.40,10.69){$\cdot$}
\put(07.05,09.11){$\cdot$} \put(07.39,10.63){$\cdot$} \put(07.49,09.25){$\cdot$} \put(07.45,10.64){$\cdot$} \put(09.22,10.44){$\cdot$} \put(07.83,09.98){$\cdot$} \put(07.50,10.56){$\cdot$}
\put(07.59,09.11){$\cdot$} \put(09.43,10.55){$\cdot$} \put(08.49,09.85){$\cdot$} \put(07.77,08.99){$\cdot$} \put(07.62,10.15){$\cdot$} \put(08.07,09.84){$\cdot$} \put(08.94,10.10){$\cdot$}
\put(08.11,09.92){$\cdot$} \put(05.08,10.18){$\cdot$} \put(07.68,09.85){$\cdot$} \put(08.03,09.81){$\cdot$} \put(09.29,09.31){$\cdot$} \put(10.21,09.33){$\cdot$} \put(07.79,09.38){$\cdot$}
\put(08.91,10.10){$\cdot$} \put(07.55,09.98){$\cdot$} \put(08.54,09.81){$\cdot$} \put(07.67,10.15){$\cdot$} \put(09.46,09.80){$\cdot$} \put(07.86,10.46){$\cdot$} \put(07.13,10.48){$\cdot$}
\put(08.10,09.12){$\cdot$} \put(08.38,09.76){$\cdot$} \put(07.90,09.26){$\cdot$} \put(08.52,10.37){$\cdot$} \put(07.22,08.94){$\cdot$} \put(04.46,09.72){$\cdot$} \put(07.99,09.79){$\cdot$}
\put(06.39,09.95){$\cdot$} \put(08.35,09.41){$\cdot$} \put(08.91,09.08){$\cdot$} \put(08.11,09.36){$\cdot$} \put(07.99,09.49){$\cdot$} \put(07.88,09.73){$\cdot$} \put(08.84,10.04){$\cdot$}
\put(07.61,09.02){$\cdot$} \put(08.98,09.50){$\cdot$} \put(07.95,10.44){$\cdot$} \put(07.94,10.16){$\cdot$} \put(04.00,10.44){$\cdot$} \put(05.50,09.94){$\cdot$} \put(08.60,09.35){$\cdot$}
\put(07.57,10.64){$\cdot$} \put(07.24,09.84){$\cdot$} \put(06.03,09.52){$\cdot$} \put(10.37,10.54){$\cdot$} \put(05.12,09.83){$\cdot$} \put(07.88,10.58){$\cdot$} \put(05.47,09.09){$\cdot$}
\put(07.99,10.32){$\cdot$} \put(10.58,10.55){$\cdot$} \put(08.13,09.40){$\cdot$} \put(08.37,10.62){$\cdot$} \put(09.33,09.36){$\cdot$} \put(10.60,10.48){$\cdot$} \put(05.43,10.33){$\cdot$}
\put(08.25,10.40){$\cdot$} \put(05.81,09.45){$\cdot$} \put(07.81,08.90){$\cdot$} \put(08.07,09.53){$\cdot$} \put(11.34,09.84){$\cdot$} \put(06.97,10.35){$\cdot$} \put(05.43,09.32){$\cdot$}
\put(08.36,08.91){$\cdot$} \put(07.72,09.94){$\cdot$} \put(08.11,09.14){$\cdot$} \put(07.59,09.45){$\cdot$} \put(06.46,09.96){$\cdot$} \put(08.15,10.18){$\cdot$} \put(07.95,09.99){$\cdot$}
\put(08.83,10.43){$\cdot$} \put(08.05,09.10){$\cdot$} \put(07.81,09.37){$\cdot$} \put(07.54,08.96){$\cdot$} \put(08.18,09.78){$\cdot$} \put(08.06,09.16){$\cdot$} \put(07.19,08.95){$\cdot$}
\put(07.80,09.01){$\cdot$} \put(07.97,09.84){$\cdot$} \put(07.83,10.49){$\cdot$} \put(09.20,08.98){$\cdot$} \put(08.92,09.28){$\cdot$} \put(08.33,09.37){$\cdot$} \put(06.63,09.95){$\cdot$}
\put(09.37,10.52){$\cdot$} \put(07.17,09.84){$\cdot$} \put(07.16,09.69){$\cdot$} \put(08.75,09.57){$\cdot$} \put(07.39,10.29){$\cdot$} \put(07.73,09.66){$\cdot$} \put(07.05,09.58){$\cdot$}
\put(09.78,09.84){$\cdot$} \put(07.96,10.60){$\cdot$} \put(08.42,09.67){$\cdot$} \put(07.58,09.72){$\cdot$} \put(08.44,09.21){$\cdot$} \put(07.19,09.18){$\cdot$} \put(08.49,10.42){$\cdot$}
\put(08.13,09.49){$\cdot$} \put(09.17,09.94){$\cdot$} \put(06.95,09.80){$\cdot$} \put(09.10,09.37){$\cdot$} \put(10.30,08.94){$\cdot$} \put(08.22,10.33){$\cdot$} \put(07.83,08.92){$\cdot$}
\put(07.54,10.31){$\cdot$} \put(06.66,09.69){$\cdot$} \put(06.34,09.38){$\cdot$} \put(07.86,10.09){$\cdot$} \put(08.08,10.69){$\cdot$} \put(08.05,09.35){$\cdot$} \put(06.54,09.35){$\cdot$}
\put(08.07,09.87){$\cdot$} \put(08.26,09.67){$\cdot$} \put(06.82,10.41){$\cdot$} \put(07.88,10.64){$\cdot$} \put(07.13,09.10){$\cdot$} \put(08.61,09.08){$\cdot$} \put(07.61,09.94){$\cdot$}
\put(05.39,09.62){$\cdot$} \put(09.90,08.92){$\cdot$} \put(07.79,09.37){$\cdot$} \put(08.22,09.30){$\cdot$} \put(09.41,09.57){$\cdot$} \put(08.71,09.02){$\cdot$} \put(07.28,10.59){$\cdot$}
\put(09.05,10.36){$\cdot$} \put(10.82,08.99){$\cdot$} \put(05.84,09.05){$\cdot$} \put(07.33,10.04){$\cdot$} \put(11.89,09.23){$\cdot$} \put(07.12,08.91){$\cdot$} \put(08.06,09.89){$\cdot$}
\put(04.77,09.57){$\cdot$} \put(07.67,10.06){$\cdot$} \put(07.77,10.16){$\cdot$} \put(09.33,09.03){$\cdot$} \put(05.01,10.57){$\cdot$} \put(07.17,10.07){$\cdot$} \put(07.31,09.32){$\cdot$}
\put(04.78,09.05){$\cdot$} \put(06.93,09.38){$\cdot$} \put(06.61,09.34){$\cdot$} \put(08.56,10.65){$\cdot$} \put(07.25,10.23){$\cdot$} \put(07.45,09.03){$\cdot$} \put(08.50,09.75){$\cdot$}
\put(06.75,10.62){$\cdot$} \put(07.92,09.92){$\cdot$} \put(07.94,10.06){$\cdot$} \put(07.58,09.74){$\cdot$} \put(06.55,09.91){$\cdot$} \put(06.47,10.36){$\cdot$} \put(09.53,10.37){$\cdot$}
\put(09.05,09.44){$\cdot$} \put(06.29,09.10){$\cdot$} \put(09.08,10.35){$\cdot$} \put(05.54,09.27){$\cdot$} \put(04.48,09.81){$\cdot$} \put(05.19,09.24){$\cdot$} \put(05.80,09.56){$\cdot$}
\put(09.39,09.24){$\cdot$} \put(07.11,09.55){$\cdot$} \put(07.79,08.95){$\cdot$} \put(07.67,10.58){$\cdot$} \put(09.69,10.10){$\cdot$} \put(08.29,10.45){$\cdot$} \put(08.31,10.10){$\cdot$}
\put(07.52,09.84){$\cdot$} \put(09.19,10.39){$\cdot$} \put(08.28,10.11){$\cdot$} \put(09.70,10.68){$\cdot$} \put(03.99,09.01){$\cdot$} \put(08.61,10.29){$\cdot$} \put(08.27,10.09){$\cdot$}

\thinlines

\put(6.5,11.5){\vector(-1,0){1}}
\put(6.5,11.25){\line(0,1){0.5}}
\put(5.5,11.9){\tiny{$C_{L \beta}$}}

\put(7.5,11.5){\vector(1,0){0.3}}
\put(7.5,11.25){\line(0,1){0.5}}
\put(7.0,11.9){\tiny{$C_{L \alpha}$}}

\put(8.5,11.5){\vector(-1,0){0.3}}
\put(8.5,11.25){\line(0,1){0.5}}
\put(8.0,11.9){\tiny{$C_{L \alpha}$}}

\put(9.5,11.5){\vector(1,0){1}}
\put(9.5,11.25){\line(0,1){0.5}}
\put(9.5,11.9){\tiny{$C_{L \beta}$}}

\put(1,8){b) stage 1: dissolution of interlayer}

\thicklines
\put(1,3){\line(1,0){14}}
\put(1,5){\line(1,0){14}}
\put(1,3){\line(0,1){2}}
\put(15,3){\line(0,1){2}}

\put(01.34,03.55){$\cdot$} \put(03.98,03.80){$\cdot$} \put(07.41,03.54){$\cdot$} \put(05.93,03.13){$\cdot$} \put(02.97,03.55){$\cdot$} \put(04.47,03.54){$\cdot$} \put(06.67,03.20){$\cdot$}
\put(10.90,03.36){$\cdot$} \put(12.18,03.67){$\cdot$} \put(11.86,04.01){$\cdot$} \put(12.55,03.72){$\cdot$} \put(06.46,04.50){$\cdot$} \put(02.02,03.38){$\cdot$} \put(07.18,03.84){$\cdot$}
\put(12.04,04.09){$\cdot$} \put(04.24,04.49){$\cdot$} \put(10.01,04.32){$\cdot$} \put(02.14,03.33){$\cdot$} \put(03.38,04.50){$\cdot$} \put(07.14,03.51){$\cdot$} \put(07.50,04.08){$\cdot$}
\put(13.81,04.36){$\cdot$} \put(09.25,03.81){$\cdot$} \put(08.30,04.41){$\cdot$} \put(11.18,04.10){$\cdot$} \put(06.70,04.08){$\cdot$} \put(03.36,03.53){$\cdot$} \put(04.89,03.31){$\cdot$}
\put(09.69,02.94){$\cdot$} \put(14.04,03.09){$\cdot$} \put(04.84,04.42){$\cdot$} \put(02.69,03.22){$\cdot$} \put(01.18,04.51){$\cdot$} \put(03.44,03.85){$\cdot$} \put(05.23,02.95){$\cdot$}
\put(04.65,03.08){$\cdot$} \put(04.57,03.82){$\cdot$} \put(06.06,03.67){$\cdot$} \put(14.37,03.89){$\cdot$} \put(13.16,02.91){$\cdot$} \put(06.40,03.96){$\cdot$} \put(12.52,03.97){$\cdot$}
\put(12.19,04.42){$\cdot$} \put(02.99,04.35){$\cdot$} \put(12.72,04.45){$\cdot$} \put(01.11,04.23){$\cdot$} \put(07.13,04.05){$\cdot$} \put(04.19,04.66){$\cdot$} \put(09.28,04.66){$\cdot$}
\put(05.42,03.52){$\cdot$} \put(04.26,03.02){$\cdot$} \put(08.99,03.13){$\cdot$} \put(05.82,04.28){$\cdot$} \put(14.18,04.30){$\cdot$} \put(05.91,04.11){$\cdot$} \put(06.49,03.17){$\cdot$}
\put(03.58,04.34){$\cdot$} \put(11.26,03.54){$\cdot$} \put(08.66,03.33){$\cdot$} \put(07.96,03.95){$\cdot$} \put(08.07,04.54){$\cdot$} \put(13.47,03.92){$\cdot$} \put(03.10,04.66){$\cdot$}
\put(09.77,03.90){$\cdot$} \put(07.30,04.53){$\cdot$} \put(06.13,03.02){$\cdot$} \put(03.58,04.07){$\cdot$} \put(14.54,03.24){$\cdot$} \put(13.33,03.83){$\cdot$} \put(09.41,03.53){$\cdot$}
\put(02.18,02.90){$\cdot$} \put(11.97,04.63){$\cdot$} \put(07.00,04.20){$\cdot$} \put(06.72,03.23){$\cdot$} \put(05.37,04.20){$\cdot$} \put(14.25,04.32){$\cdot$} \put(08.62,03.46){$\cdot$}
\put(06.89,03.28){$\cdot$} \put(05.44,04.04){$\cdot$} \put(09.20,03.62){$\cdot$} \put(11.29,04.44){$\cdot$} \put(11.74,04.69){$\cdot$} \put(08.62,03.74){$\cdot$} \put(12.77,03.96){$\cdot$}
\put(09.58,02.95){$\cdot$} \put(06.06,03.70){$\cdot$} \put(07.57,03.73){$\cdot$} \put(03.98,03.40){$\cdot$} \put(01.48,04.10){$\cdot$} \put(08.17,03.82){$\cdot$} \put(08.70,04.64){$\cdot$}
\put(06.43,03.51){$\cdot$} \put(13.83,03.75){$\cdot$} \put(03.27,03.59){$\cdot$} \put(08.36,04.09){$\cdot$} \put(10.27,04.22){$\cdot$} \put(11.26,02.93){$\cdot$} \put(08.08,03.43){$\cdot$}
\put(09.37,03.95){$\cdot$} \put(10.88,04.52){$\cdot$} \put(10.03,03.82){$\cdot$} \put(09.43,04.32){$\cdot$} \put(14.12,04.38){$\cdot$} \put(09.36,04.64){$\cdot$} \put(10.42,03.39){$\cdot$}
\put(03.61,03.22){$\cdot$} \put(10.28,03.06){$\cdot$} \put(10.19,03.00){$\cdot$} \put(14.32,04.56){$\cdot$} \put(14.36,03.17){$\cdot$} \put(01.57,04.58){$\cdot$} \put(13.95,03.06){$\cdot$}
\put(13.21,04.47){$\cdot$} \put(01.66,03.41){$\cdot$} \put(01.45,04.34){$\cdot$} \put(01.50,02.97){$\cdot$} \put(12.92,04.66){$\cdot$} \put(06.12,04.12){$\cdot$} \put(05.83,02.97){$\cdot$}
\put(13.69,03.40){$\cdot$} \put(13.16,03.59){$\cdot$} \put(05.74,03.75){$\cdot$} \put(07.49,04.51){$\cdot$} \put(06.31,03.18){$\cdot$} \put(02.57,03.30){$\cdot$} \put(06.17,03.46){$\cdot$}
\put(10.22,03.13){$\cdot$} \put(14.61,03.14){$\cdot$} \put(08.77,03.30){$\cdot$} \put(05.62,04.27){$\cdot$} \put(10.58,03.39){$\cdot$} \put(06.17,04.43){$\cdot$} \put(04.03,04.01){$\cdot$}
\put(08.61,04.39){$\cdot$} \put(03.28,04.42){$\cdot$} \put(08.30,03.47){$\cdot$} \put(09.38,04.29){$\cdot$} \put(08.01,03.99){$\cdot$} \put(09.28,04.31){$\cdot$} \put(09.10,03.91){$\cdot$}
\put(13.30,04.09){$\cdot$} \put(11.86,03.86){$\cdot$} \put(10.22,04.28){$\cdot$} \put(06.31,04.51){$\cdot$} \put(06.92,03.86){$\cdot$} \put(07.98,04.61){$\cdot$} \put(08.04,03.04){$\cdot$}
\put(12.57,04.03){$\cdot$} \put(07.00,04.12){$\cdot$} \put(01.89,04.12){$\cdot$} \put(06.63,04.66){$\cdot$} \put(08.42,03.73){$\cdot$} \put(09.00,03.63){$\cdot$} \put(07.87,03.24){$\cdot$}
\put(04.97,03.21){$\cdot$} \put(09.92,02.99){$\cdot$} \put(08.34,04.61){$\cdot$} \put(04.98,04.52){$\cdot$} \put(07.34,03.28){$\cdot$} \put(07.95,04.34){$\cdot$} \put(05.75,02.96){$\cdot$}
\put(06.19,04.53){$\cdot$} \put(08.68,03.75){$\cdot$} \put(13.66,04.46){$\cdot$} \put(05.07,04.43){$\cdot$} \put(05.88,03.08){$\cdot$} \put(13.05,03.92){$\cdot$} \put(07.88,04.63){$\cdot$}
\put(10.16,04.61){$\cdot$} \put(06.73,04.51){$\cdot$} \put(05.39,03.75){$\cdot$} \put(08.34,03.94){$\cdot$} \put(08.03,04.43){$\cdot$} \put(09.89,04.11){$\cdot$} \put(08.77,04.34){$\cdot$}
\put(07.79,04.39){$\cdot$} \put(05.22,04.19){$\cdot$} \put(06.81,04.30){$\cdot$} \put(07.65,04.04){$\cdot$} \put(12.11,04.64){$\cdot$} \put(06.83,02.92){$\cdot$} \put(08.12,03.75){$\cdot$}
\put(03.82,02.90){$\cdot$} \put(04.08,04.33){$\cdot$} \put(08.88,03.50){$\cdot$} \put(06.15,03.82){$\cdot$} \put(02.87,04.16){$\cdot$} \put(08.31,03.83){$\cdot$} \put(07.89,03.40){$\cdot$}
\put(06.88,03.62){$\cdot$} \put(08.20,04.56){$\cdot$} \put(07.69,04.08){$\cdot$} \put(06.26,04.61){$\cdot$} \put(04.10,04.31){$\cdot$} \put(07.15,04.55){$\cdot$} \put(06.29,04.52){$\cdot$}
\put(10.69,04.18){$\cdot$} \put(06.92,03.42){$\cdot$} \put(03.80,02.95){$\cdot$} \put(07.89,03.79){$\cdot$} \put(08.25,03.79){$\cdot$} \put(06.82,03.08){$\cdot$} \put(09.58,03.01){$\cdot$}
\put(07.95,04.36){$\cdot$} \put(07.01,04.39){$\cdot$} \put(08.94,02.95){$\cdot$} \put(10.93,03.36){$\cdot$} \put(04.91,03.18){$\cdot$} \put(09.68,03.08){$\cdot$} \put(06.74,02.90){$\cdot$}
\put(05.50,04.70){$\cdot$} \put(06.60,03.60){$\cdot$} \put(04.37,03.32){$\cdot$} \put(07.19,03.94){$\cdot$} \put(07.29,04.70){$\cdot$} \put(08.91,04.55){$\cdot$} \put(12.26,04.64){$\cdot$}
\put(05.70,04.26){$\cdot$} \put(08.41,02.94){$\cdot$} \put(08.48,04.43){$\cdot$} \put(11.42,04.43){$\cdot$} \put(07.60,03.32){$\cdot$} \put(07.91,03.48){$\cdot$} \put(07.40,03.46){$\cdot$}
\put(10.62,02.90){$\cdot$} \put(07.70,04.33){$\cdot$} \put(08.18,03.78){$\cdot$} \put(05.91,04.15){$\cdot$} \put(10.94,04.20){$\cdot$} \put(11.10,03.05){$\cdot$} \put(08.08,03.28){$\cdot$}
\put(06.31,03.33){$\cdot$} \put(07.75,03.78){$\cdot$} \put(12.45,04.17){$\cdot$} \put(09.47,04.28){$\cdot$} \put(04.97,03.93){$\cdot$} \put(05.78,03.07){$\cdot$} \put(08.44,04.20){$\cdot$}
\put(02.91,03.13){$\cdot$} \put(07.34,03.17){$\cdot$} \put(07.72,03.06){$\cdot$} \put(08.06,03.17){$\cdot$} \put(09.49,03.98){$\cdot$} \put(10.05,03.15){$\cdot$} \put(08.82,03.10){$\cdot$}
\put(09.00,03.82){$\cdot$} \put(03.62,03.39){$\cdot$} \put(07.99,04.19){$\cdot$} \put(07.54,03.08){$\cdot$} \put(08.94,04.63){$\cdot$} \put(07.72,03.37){$\cdot$} \put(08.28,04.68){$\cdot$}
\put(05.39,03.58){$\cdot$} \put(08.01,04.08){$\cdot$} \put(02.69,04.69){$\cdot$} \put(03.22,04.46){$\cdot$} \put(05.95,04.19){$\cdot$} \put(06.76,04.29){$\cdot$} \put(07.61,02.98){$\cdot$}
\put(06.85,04.06){$\cdot$} \put(07.39,02.96){$\cdot$} \put(10.81,03.29){$\cdot$} \put(07.87,03.78){$\cdot$} \put(06.07,04.17){$\cdot$} \put(06.75,03.78){$\cdot$} \put(05.18,03.44){$\cdot$}

\thinlines
\put(2,5.25){solid}
\put(7.15,5.25){liquid}
\put(12,5.25){solid}
\put(6.0,5.5){\vector(-1,0){1}}
\put(6.0,5.25){\line(0,1){0.5}}
\put(5.8,5.9){\tiny{$C_{L \beta}$}}
\put(10,5.5){\vector(1,0){1}}
\put(10,5.25){\line(0,1){0.5}}
\put(9.8,5.9){\tiny{$C_{L \beta}$}}
\put(1,2){c) stage 2: widening of the liquid phase}

\end{picture}\\
Fig. 4: transient liquid phase bonding
\end{center}

\newpage
\onecolumngrid
~

~

\smallskip
\twocolumngrid

\noindent
During the third stage additives will continue to diffuse out of the
liquid into B but this solid state diffusion is a much slower process.
The border of the liquid now moves inwards as the concentration of the
additive falls below $C_{L \beta}$ (figure 4d). After typically
some hours the joint finally solidifies completely. Only a small
concentration gradient remains.

\bigskip
\noindent
After this bonding process there is a long period of solid state
homogenization through diffusion. It may happen at bonding temperature
or at room temperature and takes place on a long timescale. In practice
the four stages of the process are not that ideally separated as presented
here.

\bigskip
\begin{center}
\unitlength0.5cm
\begin{picture}(16,8)

\thicklines
\put(1,3){\line(1,0){14}}
\put(1,5){\line(1,0){14}}
\put(1,3){\line(0,1){2}}
\put(15,3){\line(0,1){2}}

\put(01.34,03.55){$\cdot$} \put(03.98,03.80){$\cdot$} \put(07.41,03.54){$\cdot$} \put(05.93,03.13){$\cdot$} \put(02.97,03.55){$\cdot$} \put(04.47,03.54){$\cdot$} \put(06.67,03.20){$\cdot$}
\put(10.90,03.36){$\cdot$} \put(12.18,03.67){$\cdot$} \put(11.86,04.01){$\cdot$} \put(12.55,03.72){$\cdot$} \put(06.46,04.50){$\cdot$} \put(02.02,03.38){$\cdot$} \put(07.18,03.84){$\cdot$}
\put(12.04,04.09){$\cdot$} \put(04.24,04.49){$\cdot$} \put(10.01,04.32){$\cdot$} \put(02.14,03.33){$\cdot$} \put(03.38,04.50){$\cdot$} \put(07.14,03.51){$\cdot$} \put(07.50,04.08){$\cdot$}
\put(13.81,04.36){$\cdot$} \put(09.25,03.81){$\cdot$} \put(08.30,04.41){$\cdot$} \put(11.18,04.10){$\cdot$} \put(06.70,04.08){$\cdot$} \put(03.36,03.53){$\cdot$} \put(04.89,03.31){$\cdot$}
\put(09.69,02.94){$\cdot$} \put(14.04,03.09){$\cdot$} \put(04.84,04.42){$\cdot$} \put(02.69,03.22){$\cdot$} \put(01.18,04.51){$\cdot$} \put(03.44,03.85){$\cdot$} \put(05.23,02.95){$\cdot$}
\put(04.65,03.08){$\cdot$} \put(04.57,03.82){$\cdot$} \put(06.06,03.67){$\cdot$} \put(14.37,03.89){$\cdot$} \put(13.16,02.91){$\cdot$} \put(06.40,03.96){$\cdot$} \put(12.52,03.97){$\cdot$}
\put(12.19,04.42){$\cdot$} \put(02.99,04.35){$\cdot$} \put(12.72,04.45){$\cdot$} \put(01.11,04.23){$\cdot$} \put(07.13,04.05){$\cdot$} \put(04.19,04.66){$\cdot$} \put(09.28,04.66){$\cdot$}
\put(05.42,03.52){$\cdot$} \put(04.26,03.02){$\cdot$} \put(08.99,03.13){$\cdot$} \put(05.82,04.28){$\cdot$} \put(14.18,04.30){$\cdot$} \put(05.91,04.11){$\cdot$} \put(06.49,03.17){$\cdot$}
\put(03.58,04.34){$\cdot$} \put(11.26,03.54){$\cdot$} \put(08.66,03.33){$\cdot$} \put(07.96,03.95){$\cdot$} \put(08.07,04.54){$\cdot$} \put(13.47,03.92){$\cdot$} \put(03.10,04.66){$\cdot$}
\put(09.77,03.90){$\cdot$} \put(07.30,04.53){$\cdot$} \put(06.13,03.02){$\cdot$} \put(03.58,04.07){$\cdot$} \put(14.54,03.24){$\cdot$} \put(13.33,03.83){$\cdot$} \put(09.41,03.53){$\cdot$}
\put(02.18,02.90){$\cdot$} \put(11.97,04.63){$\cdot$} \put(07.00,04.20){$\cdot$} \put(06.72,03.23){$\cdot$} \put(05.37,04.20){$\cdot$} \put(14.25,04.32){$\cdot$} \put(08.62,03.46){$\cdot$}
\put(06.89,03.28){$\cdot$} \put(05.44,04.04){$\cdot$} \put(09.20,03.62){$\cdot$} \put(11.29,04.44){$\cdot$} \put(11.74,04.69){$\cdot$} \put(08.62,03.74){$\cdot$} \put(12.77,03.96){$\cdot$}
\put(09.58,02.95){$\cdot$} \put(06.06,03.70){$\cdot$} \put(07.57,03.73){$\cdot$} \put(03.98,03.40){$\cdot$} \put(01.48,04.10){$\cdot$} \put(08.17,03.82){$\cdot$} \put(08.70,04.64){$\cdot$}
\put(06.43,03.51){$\cdot$} \put(13.83,03.75){$\cdot$} \put(03.27,03.59){$\cdot$} \put(08.36,04.09){$\cdot$} \put(10.27,04.22){$\cdot$} \put(11.26,02.93){$\cdot$} \put(08.08,03.43){$\cdot$}
\put(09.37,03.95){$\cdot$} \put(10.88,04.52){$\cdot$} \put(10.03,03.82){$\cdot$} \put(09.43,04.32){$\cdot$} \put(14.12,04.38){$\cdot$} \put(09.36,04.64){$\cdot$} \put(10.42,03.39){$\cdot$}
\put(03.61,03.22){$\cdot$} \put(10.28,03.06){$\cdot$} \put(10.19,03.00){$\cdot$} \put(14.32,04.56){$\cdot$} \put(14.36,03.17){$\cdot$} \put(01.57,04.58){$\cdot$} \put(13.95,03.06){$\cdot$}
\put(13.21,04.47){$\cdot$} \put(01.66,03.41){$\cdot$} \put(01.45,04.34){$\cdot$} \put(01.50,02.97){$\cdot$} \put(12.92,04.66){$\cdot$} \put(06.12,04.12){$\cdot$} \put(05.83,02.97){$\cdot$}
\put(13.69,03.40){$\cdot$} \put(13.16,03.59){$\cdot$} \put(05.74,03.75){$\cdot$} \put(07.49,04.51){$\cdot$} \put(06.31,03.18){$\cdot$} \put(02.57,03.30){$\cdot$} \put(06.17,03.46){$\cdot$}
\put(10.22,03.13){$\cdot$} \put(14.61,03.14){$\cdot$} \put(08.77,03.30){$\cdot$} \put(05.62,04.27){$\cdot$} \put(10.58,03.39){$\cdot$} \put(06.17,04.43){$\cdot$} \put(04.03,04.01){$\cdot$}
\put(08.61,04.39){$\cdot$} \put(03.28,04.42){$\cdot$} \put(08.30,03.47){$\cdot$} \put(09.38,04.29){$\cdot$} \put(08.01,03.99){$\cdot$} \put(09.28,04.31){$\cdot$} \put(09.10,03.91){$\cdot$}
\put(13.30,04.09){$\cdot$} \put(11.86,03.86){$\cdot$} \put(10.22,04.28){$\cdot$} \put(06.31,04.51){$\cdot$} \put(06.92,03.86){$\cdot$} \put(07.98,04.61){$\cdot$} \put(08.04,03.04){$\cdot$}
\put(12.57,04.03){$\cdot$} \put(07.00,04.12){$\cdot$} \put(01.89,04.12){$\cdot$} \put(06.63,04.66){$\cdot$} \put(08.42,03.73){$\cdot$} \put(09.00,03.63){$\cdot$} \put(07.87,03.24){$\cdot$}
\put(04.97,03.21){$\cdot$} \put(09.92,02.99){$\cdot$} \put(08.34,04.61){$\cdot$} \put(04.98,04.52){$\cdot$} \put(07.34,03.28){$\cdot$} \put(07.95,04.34){$\cdot$} \put(05.75,02.96){$\cdot$}
\put(06.19,04.53){$\cdot$} \put(08.68,03.75){$\cdot$} \put(13.66,04.46){$\cdot$} \put(05.07,04.43){$\cdot$} \put(05.88,03.08){$\cdot$} \put(13.05,03.92){$\cdot$} \put(07.88,04.63){$\cdot$}
\put(10.16,04.61){$\cdot$} \put(06.73,04.51){$\cdot$} \put(05.39,03.75){$\cdot$} \put(08.34,03.94){$\cdot$} \put(08.03,04.43){$\cdot$} \put(09.89,04.11){$\cdot$} \put(08.77,04.34){$\cdot$}
\put(07.79,04.39){$\cdot$} \put(05.22,04.19){$\cdot$} \put(06.81,04.30){$\cdot$} \put(07.65,04.04){$\cdot$} \put(12.11,04.64){$\cdot$} \put(06.83,02.92){$\cdot$} \put(08.12,03.75){$\cdot$}
\put(03.82,02.90){$\cdot$} \put(04.08,04.33){$\cdot$} \put(08.88,03.50){$\cdot$} \put(06.15,03.82){$\cdot$} \put(02.87,04.16){$\cdot$} \put(08.31,03.83){$\cdot$} \put(07.89,03.40){$\cdot$}
\put(06.88,03.62){$\cdot$} \put(08.20,04.56){$\cdot$} \put(07.69,04.08){$\cdot$} \put(06.26,04.61){$\cdot$} \put(04.10,04.31){$\cdot$} \put(07.15,04.55){$\cdot$} \put(06.29,04.52){$\cdot$}
\put(10.69,04.18){$\cdot$} \put(06.92,03.42){$\cdot$} \put(03.80,02.95){$\cdot$} \put(07.89,03.79){$\cdot$} \put(08.25,03.79){$\cdot$} \put(06.82,03.08){$\cdot$} \put(09.58,03.01){$\cdot$}
\put(07.95,04.36){$\cdot$} \put(07.01,04.39){$\cdot$} \put(08.94,02.95){$\cdot$} \put(10.93,03.36){$\cdot$} \put(04.91,03.18){$\cdot$} \put(09.68,03.08){$\cdot$} \put(06.74,02.90){$\cdot$}
\put(05.50,04.70){$\cdot$} \put(06.60,03.60){$\cdot$} \put(04.37,03.32){$\cdot$} \put(07.19,03.94){$\cdot$} \put(07.29,04.70){$\cdot$} \put(08.91,04.55){$\cdot$} \put(12.26,04.64){$\cdot$}
\put(05.70,04.26){$\cdot$} \put(08.41,02.94){$\cdot$} \put(08.48,04.43){$\cdot$} \put(11.42,04.43){$\cdot$} \put(07.60,03.32){$\cdot$} \put(07.91,03.48){$\cdot$} \put(07.40,03.46){$\cdot$}
\put(10.62,02.90){$\cdot$} \put(07.70,04.33){$\cdot$} \put(08.18,03.78){$\cdot$} \put(05.91,04.15){$\cdot$} \put(10.94,04.20){$\cdot$} \put(11.10,03.05){$\cdot$} \put(08.08,03.28){$\cdot$}
\put(06.31,03.33){$\cdot$} \put(07.75,03.78){$\cdot$} \put(12.45,04.17){$\cdot$} \put(09.47,04.28){$\cdot$} \put(04.97,03.93){$\cdot$} \put(05.78,03.07){$\cdot$} \put(08.44,04.20){$\cdot$}
\put(02.91,03.13){$\cdot$} \put(07.34,03.17){$\cdot$} \put(07.72,03.06){$\cdot$} \put(08.06,03.17){$\cdot$} \put(09.49,03.98){$\cdot$} \put(10.05,03.15){$\cdot$} \put(08.82,03.10){$\cdot$}
\put(09.00,03.82){$\cdot$} \put(03.62,03.39){$\cdot$} \put(07.99,04.19){$\cdot$} \put(07.54,03.08){$\cdot$} \put(08.94,04.63){$\cdot$} \put(07.72,03.37){$\cdot$} \put(08.28,04.68){$\cdot$}
\put(05.39,03.58){$\cdot$} \put(08.01,04.08){$\cdot$} \put(02.69,04.69){$\cdot$} \put(03.22,04.46){$\cdot$} \put(05.95,04.19){$\cdot$} \put(06.76,04.29){$\cdot$} \put(07.61,02.98){$\cdot$}
\put(06.85,04.06){$\cdot$} \put(07.39,02.96){$\cdot$} \put(10.81,03.29){$\cdot$} \put(07.87,03.78){$\cdot$} \put(06.07,04.17){$\cdot$} \put(06.75,03.78){$\cdot$} \put(05.18,03.44){$\cdot$}

\thinlines
\put(2,5.25){solid}
\put(7.15,5.25){liquid}
\put(12,5.25){solid}
\put(6.0,5.5){\vector(1,0){1}}
\put(6.0,5.25){\line(0,1){0.5}}
\put(5.8,5.9){\tiny{$C_{L \beta}$}}
\put(10,5.5){\vector(-1,0){1}}
\put(10,5.25){\line(0,1){0.5}}
\put(9.8,5.9){\tiny{$C_{L \beta}$}}
\put(1,2){d) stage 3: isothermal solidification}

\end{picture}\\
Fig. 4: transient liquid phase bonding\\
(continued)
\end{center}

\onecolumngrid

~

\bigskip
\noindent
~

\section{III. Sample Preparation}

\noindent
A sample featuring a TLP bond was provided by The Welding Institute in Abington, UK.
The sample under investigation was produced as part of a trial to optimise the process parameters of TLP bonding
in terms of brazing temperature and time at temperature.
This sample had been heated to $1150^\circ C$ which is below the recommended brazing temperature for this foil ($1195^\circ C$).
As a result the sample features a frozen unfinished joint.

\bigskip
\noindent
The foil used for bonding was a grade MBF-51.
Its chemical composition is $15\%$ Cr, $7.25\%$ Si, $1.4\%$ B, $0.06\%$ C (in $wt.\%$) and the balance is Ni. Both
the Si and the B are the melting point depressants - they are both about as effective as each other as temperature depressants.
MBF-51 is an amorphous foil - its liquidus is $1126^\circ C$ - its solidus is $1030^\circ C$ and the recommended braze temperature is $1195^\circ C$.
The base material of the specimen was steel.
The grade of steel was a 50D (S355J2G3) with a chemical composition of $0.22\%$ C, $0.55\%$ Si, $1.6\%$ Mn, $0.035\%$ P and $0.035\%$ S.

\bigskip
\noindent
A neutron beam penetrating the sample of thickness $\Delta z=0.5cm$ is attenuated by capture and incoherent scattering according to
$- \log(I/I_0) = \mu \cdot \Delta z$. The expected attenuation coefficient for steel is $\mu_{steel} = 1.20 cm^{-1}$. The
attenuation is dominated by the capture and scattering on iron ($\mu_{Fe,sc} = 0.96 cm^{-1}$, $\mu_{Fe,ab} = 0.21 cm^{-1}$) which together ammount for $98\%$ of
the attenuation. The foil has a higher attenuation of $\mu_{foil} \approx 7 cm^{-1}$, dominated by capture in boron ($\mu_{B,ab} = 5.3 cm^{-1}$)
and scattering on Nickel ($\mu_{Ni,sc}=1.3 cm^{-1}$).

\bigskip
\noindent
The bonding profile consisted of heating from room
temperature to $1000^\circ C$ at $7.5^\circ C\,min^{-1}$ and dwell for $5 min.$, $1000^\circ C$ to $1150^\circ C$ at $5^\circ C\,min^{-1}$ and dwell 
\twocolumngrid

\noindent
for $15 min.$,
and of cooling from $1150^\circ C$ to room temperature at $5^\circ C\,min^{-1}$. 
A load of $5\,kN$ was applied on the heating cycle at $1000^\circ C$ and removed when the temperature fell
below $1000^\circ C$ on the cooling cycle. The whole process was carried out in a vacuum at $10^{-4} mbar$.

\bigskip
\noindent
After the center TLP bond was finished, two cut-offs have been made on the right and on the left
of the sample using a spark
erosion machine. Their 
surfaces are level to about $15\mu m \pm 5\mu m$. 
The left gap contains a bonding foil
and the right gap is empty.

\bigskip
\noindent
\begin{center}
\epsfig{file=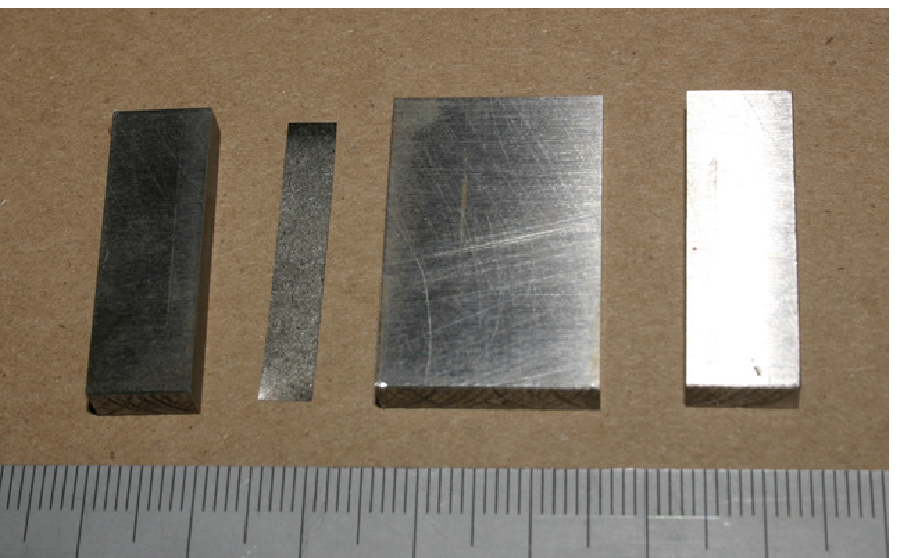,width=5.5cm}\\
Fig. 5: preparation of the sample
\\(l. to r.: steel, bonding foil,\\
steel containing TLP bond, steel)

\end{center}

\onecolumngrid

\newpage
\section{IV. Experimental Setup}

\noindent
The neutron radiography station is located at the high-flux research reactor of the
Institut Laue-Langevin in Grenoble, France \cite{schneider, neutrograph}. 
Free neutrons are produced in a fission reactor and enter the experimental area
through a free flight tube. In a distance of $15m$ from the reactor core there is a thermal neutron flux of
$3 \cdot 10^9\;s^{-1}\,cm^{-2}$. The beam is about $16cm$ by $18cm$
wide and has a divergence of less than $0.5^\circ$. For the purpose of this paper the beam may be treated as perfectly parallel.

\bigskip
\noindent
The sample under investigation is placed in front of the beam window
and attenuates the neutron beam. While X-rays interact electromagnetically
with the electron shells in the atoms neutrons only interact strongly
upon collision with a nucleus. Therefore the cross-section of neutrons
does not rise with the fourth power of the element number as in the case of X-rays but rather shows an irregular pattern depending on the scattering
or capturing isotope. This is why many heavy elements such as metals
which are opaque to X-rays can be easily penetrated with neutrons.

\bigskip
\noindent
For the investigation of the TLP joint use is made of the fact that the
bonding interlayer contains $1.4\%$ of boron which heavily absorbs neutrons \cite{NRboron}.

\bigskip
\noindent
The neutrons which are neither scattered out of the beam nor absorbed
in the sample hit a detector. On a scintillator they produce light via
a ($n$,$\alpha$)-reaction. The image is captured through mirrors and
optics by a sensitive charge coupled device (CCD)-camera \cite{NRcam1}.
The resulting two-dimensional greyscale map is a representation of the
density distribution of absorbing and scattering nuclei inside the sample.

\bigskip
\begin{center}

\unitlength0.5cm
\begin{picture}(24,20)
\thicklines


\put(5.5,12.5){\epsfig{file=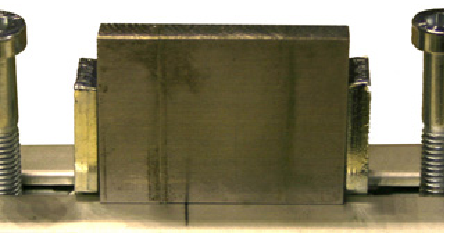,width=3cm}}

\put(2.5,7){
\put(-0.5,-0.5){\line(1,0){7}}
\put(-0.5,-0.5){\line(0,1){5}}
\put(-0.5,4.5){\line(1,0){7}}
\put(6.5,-0.5){\line(0,1){5}}
\epsfig{file=attenuation.eps,width=3cm}
}

\thinlines

\put(7.5,18.5){\vector(-2,-3){1}}
\put(9.5,18.5){\vector(-2,-3){1}}
\put(13.5,18.5){\vector(-2,-3){1}}
\put(8,19){incident neutrons}

\put(8.2,12){transmitted neutrons}

\put(13,14){sample}

\put(10,8.75){scintillator}

\put(4,5.5){converted photons}

\put(4.5,1){Al-coated Si-mirrors}

\put(20,15){CCD-camera}

\multiput(2,4)(2,3){5}{\line(2,3){.5}}
\multiput(3,5.5)(2,3){5}{\line(2,3){.5}}

\multiput(2,4)(2,0){8}{\line(1,0){1}}

\put(17,4){\line(2,3){.5}}
\put(18,5.5){\line(2,3){.5}}
\put(19,7){\line(2,3){.5}}
\put(20,8.5){\vector(2,3){.5}}

\put(0,3){\line(2,-1){4}}
\put(0,3){\line(0,1){4}}
\put(0,7){\line(2,-1){4}}
\put(4,1){\line(0,1){4}}

\put(15,1){\line(2,1){4}}
\put(15,1){\line(0,1){4}}
\put(15,5){\line(2,1){4}}
\put(19,3){\line(0,1){4}}

\put(20,9){\line(0,1){2}}
\put(20,9){\line(1,0){2}}
\put(22,9){\line(0,1){2}}
\put(20,11){\line(1,0){2}}
\put(22,9){\line(2,3){2}}
\put(20,11){\line(2,3){2}}
\put(22,11){\line(2,3){2}}
\put(24,12){\line(0,1){2}}
\put(22,14){\line(1,0){2}}
\put(21,10){\circle{1}}
\put(20.8,9.7){\circle{1}}
\put(20.8,9.7){\circle{0.8}}

\put(1,15){\vector(1,0){2}}		\put(3.2,14.8){$x$}
\put(1,15){\vector(0,1){2}}		\put(.85,17.3){$y$}
\put(1,15){\vector(-2,-3){1}}	\put(-.3,13){$z$}

\end{picture}

Fig. 6: neutron radiography setup

\end{center}

\bigskip
\noindent
The attenuation can be calculated from the decrease in intensity
according to the law of exponential attenuation $I=I_0 \cdot \exp{-\int \mu \, ds}$.
Here we assume that $\mu$ is independent of $z$. The resulting map $\mu(x,y)$ is
the neutron radiography image of the sample. In case of non-flat specimen, neutron
tomography can also reveal the complete three-dimensional attenuation distribution.

\onecolumngrid

\newpage
\section{V. Data Treatment and Corrections}

\noindent
Figure 2 shows the neutron radiography of a stainless steel sample.
The sample is 39mm by 30mm in size and is clamped to a sample holder.

\bigskip
\twocolumngrid

\noindent
The image has been corrected for the inhomogenity of the neutron beam intensity,
the inhomogenous detector sensitivities and detector offset.

\bigskip
\noindent
For the analysis the rectangular region indicated by the crosses was selected.
On a different scale of greylevels three vertical lines become visible in figure 7.
These lines indicate the location of three cut-offs:

\bigskip
\noindent
The center cut contains a transient liquid phase bond.
Before heat treatment the interlayer foil was $50\mu m$ thick.
The increase in neutron attenuation compared to that of the steel is almost completely due to its boron content.

\bigskip
\noindent
The left and right gap provide the reference values for the bonding foil
in its initial state and for air in comparison to the diffusion profile.

\bigskip
\noindent
Any pixel in these images is about $110 \mu m$ by $110 \mu m$ in size.
This resolution is limited by the natural divergence of the neutron
beam, the inherent resolution of the scintillator and the optical system.
The greylevels are known within a relative error of about $\Delta c / c \approx 6 \cdot 10^{-4}$
per pixel. For this precision the images were summed from $1000$ pictures of each $200 \mu s$
exposure time.

\bigskip
\noindent
By summing the density distribution $\int \mu(x,y,z) \, dz$ along the $y$-axis perpendicular to the neutron beam and acknowledging the implicit
projection along the $z$-axis an attenuation profile $\mu(x)$ is created, see figure 8.

\bigskip
\noindent
It is necessary to correct for some scattering effects.
Scattered neutrons from all over the sample contribute to the 
brightness of the image near the centre; however scattered neutrons only 
from one half space of the sample contribute to the brightness of the 
border. That is why a homogenous sample with significant scattering
looks brighter in the centre than at its borders.

\bigskip
\noindent
The detection of scattered neutrons also decreases the overall observed
attenuation, which is considerably lower than the anticipated $\mu_{steel}=1.2 \, cm^{-1}$.
Both corrections, for the gradient and the overall shift, 
yield the corrected attenuation profile shown in figure 9.

\bigskip
\noindent
~

\bigskip
\noindent
~

\bigskip
\noindent
\begin{center}
\unitlength0.5cm
\begin{picture}(14,10)
\put(0,0){\epsfig{file=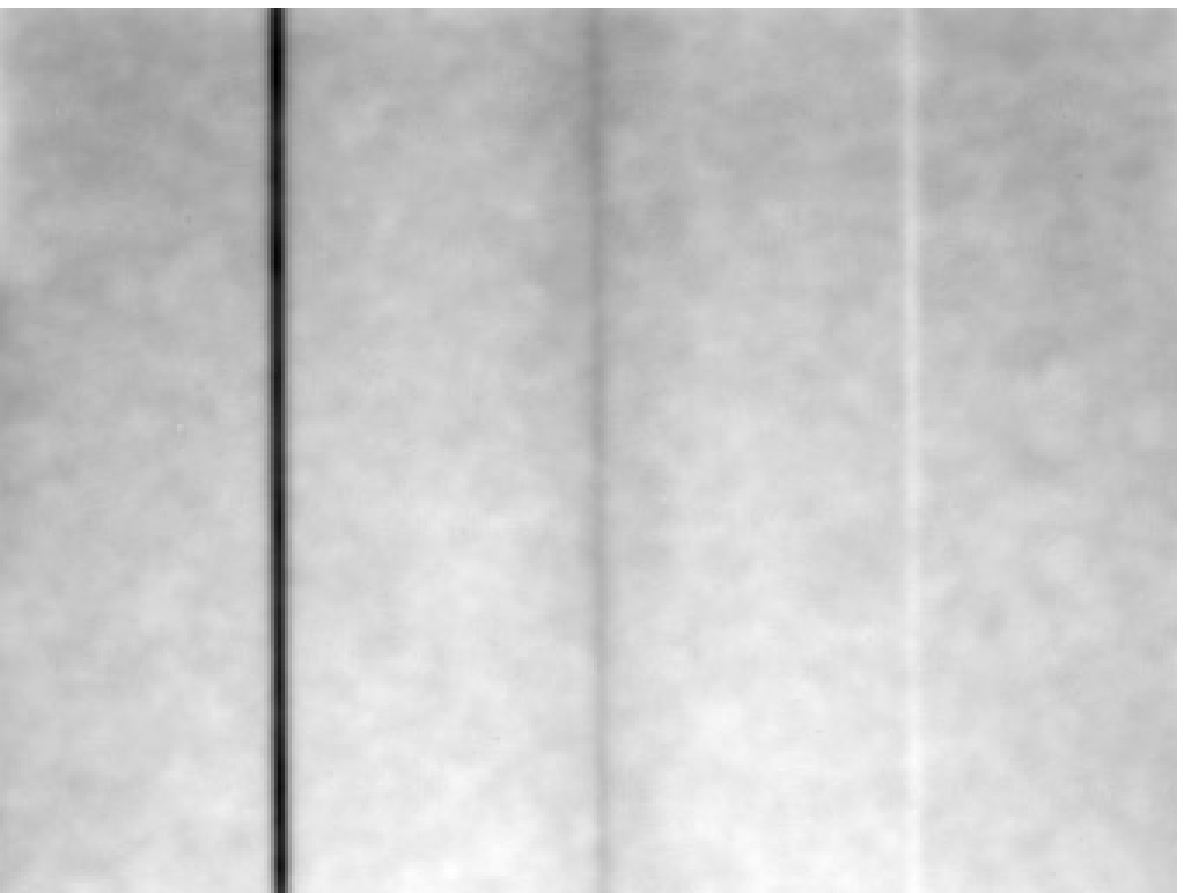,width=7cm}}

\thinlines
\put(3.5,9.7){gap with bonding}
\put(3.5,8.9){foil in initial state}
\put(3.5,8.3){/}

\put(5.5,7.5){TLP joint}
\put(7.3,6.9){/}

\put(9.2,6.5){empty gap}
\put(11.0,5.9){/}
\end{picture}\\
Fig. 7: region of interest
\end{center}

\bigskip
\begin{center}

\unitlength0.5cm
\begin{picture}(16,10)
\thicklines


\thinlines

\put(0,1){
\curve(
1.0000,2.4666, 1.0412,2.4808, 1.0823,2.4824, 1.1235,2.4749, 1.1647,2.4787, 1.2058,2.4776, 1.2470,2.4717, 1.2881,2.4680, 
1.3293,2.4605, 1.3705,2.4503, 1.4116,2.4393, 1.4528,2.4347, 1.4940,2.4354, 1.5351,2.4395, 1.5763,2.4455, 1.6175,2.4507, 
1.6586,2.4528, 1.6998,2.4599, 1.7409,2.4622, 1.7821,2.4703, 1.8233,2.4718, 1.8644,2.4662, 1.9056,2.4592, 1.9468,2.4490, 
1.9879,2.4380, 2.0291,2.4338, 2.0703,2.4278, 2.1114,2.4192, 2.1526,2.4158, 2.1937,2.4083, 2.2349,2.4009, 2.2761,2.3909, 
2.3172,2.3829, 2.3584,2.3795, 2.3996,2.3744, 2.4407,2.3678, 2.4819,2.3641, 2.5231,2.3547, 2.5642,2.3529, 2.6054,2.3414, 
2.6465,2.3313, 2.6877,2.3215, 2.7289,2.3016, 2.7700,2.2834, 2.8112,2.2700, 2.8524,2.2639, 2.8935,2.2652, 2.9347,2.2561, 
2.9759,2.2460, 3.0170,2.2408, 3.0582,2.2327, 3.0994,2.2229, 3.1405,2.2250, 3.1817,2.2252, 3.2228,2.2196, 3.2640,2.2190, 
3.3052,2.2154, 3.3463,2.2031, 3.3875,2.1939, 3.4287,2.1797, 3.4698,2.1693, 3.5110,2.1646, 3.5522,2.1548, 3.5933,2.1556, 
3.6345,2.1528, 3.6756,2.1511, 3.7168,2.1508, 3.7580,2.1541, 3.7991,2.1557, 3.8403,2.1610, 3.8815,2.1664, 3.9226,2.1757, 
3.9638,2.1937, 4.0050,2.2300, 4.0461,2.2979, 4.0873,2.4545, 4.1284,2.8095, 4.1696,3.6061, 4.2108,4.9416, 4.2519,6.2198, 
4.2931,6.5000, 4.3343,5.5468, 4.3754,4.1461, 4.4166,3.0933, 4.4578,2.5456, 4.4989,2.2974, 4.5401,2.1823, 4.5812,2.1177, 
4.6224,2.0814, 4.6636,2.0631, 4.7047,2.0474, 4.7459,2.0261, 4.7871,2.0099, 4.8282,1.9976, 4.8694,1.9987, 4.9106,1.9931, 
4.9517,1.9859, 4.9929,1.9777, 5.0340,1.9697, 5.0752,1.9583, 5.1164,1.9464, 5.1575,1.9403, 5.1987,1.9308, 5.2399,1.9242, 
5.2810,1.9113, 5.3222,1.9028, 5.3634,1.8945, 5.4045,1.8783, 5.4457,1.8749, 5.4868,1.8718, 5.5280,1.8718, 5.5692,1.8693, 
5.6103,1.8679, 5.6515,1.8679, 5.6927,1.8591, 5.7338,1.8511, 5.7750,1.8382, 5.8162,1.8265, 5.8573,1.8197, 5.8985,1.8120, 
5.9396,1.8056, 5.9808,1.8009, 6.0220,1.7937, 6.0631,1.7933, 6.1043,1.7952, 6.1455,1.7930, 6.1866,1.7907, 6.2278,1.7878, 
6.2690,1.7798, 6.3101,1.7799, 6.3513,1.7769, 6.3924,1.7730, 6.4336,1.7733, 6.4748,1.7765, 6.5159,1.7754, 6.5571,1.7833, 
6.5983,1.7921, 6.6394,1.8010, 6.6806,1.8119, 6.7218,1.8148, 6.7629,1.8161, 6.8041,1.8136, 6.8452,1.8282, 6.8864,1.8359, 
6.9276,1.8406, 6.9687,1.8488, 7.0099,1.8636, 7.0511,1.8774, 7.0922,1.8941, 7.1334,1.9149, 7.1746,1.9310, 7.2157,1.9497, 
7.2569,1.9753, 7.2981,1.9945, 7.3392,2.0143, 7.3804,2.0393, 7.4215,2.0631, 7.4627,2.0911, 7.5039,2.1217, 7.5450,2.1530, 
7.5862,2.1737, 7.6274,2.1950, 7.6685,2.2128, 7.7097,2.2280, 7.7509,2.2358, 7.7920,2.2571, 7.8332,2.2873, 7.8743,2.3394, 
7.9155,2.4021, 7.9567,2.4889, 7.9978,2.6044, 8.0390,2.7274, 8.0802,2.7853, 8.1213,2.7473, 8.1625,2.6300, 8.2037,2.5052, 
8.2448,2.4069, 8.2860,2.3293, 8.3271,2.2701, 8.3683,2.2326, 8.4095,2.2124, 8.4506,2.2006, 8.4918,2.1899, 8.5330,2.1813, 
8.5741,2.1730, 8.6153,2.1596, 8.6565,2.1438, 8.6976,2.1266, 8.7388,2.1038, 8.7799,2.0835, 8.8211,2.0546, 8.8623,2.0277, 
8.9034,1.9982, 8.9446,1.9696, 8.9858,1.9477, 9.0269,1.9239, 9.0681,1.9037, 9.1093,1.8780, 9.1504,1.8576, 9.1916,1.8447, 
9.2327,1.8287, 9.2739,1.8088, 9.3151,1.7944, 9.3562,1.7885, 9.3974,1.7852, 9.4386,1.7850, 9.4797,1.7887, 9.5209,1.7953, 
9.5621,1.7869, 9.6032,1.7829, 9.6444,1.7780, 9.6855,1.7699, 9.7267,1.7641, 9.7679,1.7586, 9.8090,1.7574, 9.8502,1.7500, 
9.8914,1.7576, 9.9325,1.7525, 9.9737,1.7537, 10.0149,1.7555, 10.0560,1.7538, 10.0972,1.7612, 10.1383,1.7694, 10.1795,1.7735, 
10.2207,1.7749, 10.2618,1.7833, 10.3030,1.7855, 10.3442,1.7865, 10.3853,1.7923, 10.4265,1.7945, 10.4677,1.8009, 10.5088,1.8039, 
10.5500,1.8091, 10.5911,1.8138, 10.6323,1.8172, 10.6735,1.8234, 10.7146,1.8292, 10.7558,1.8390, 10.7970,1.8415, 10.8381,1.8555, 
10.8793,1.8575, 10.9205,1.8576, 10.9616,1.8616, 11.0028,1.8648, 11.0439,1.8690, 11.0851,1.8765, 11.1263,1.8803, 11.1674,1.8905, 
11.2086,1.8956, 11.2498,1.9022, 11.2909,1.9004, 11.3321,1.9023, 11.3733,1.9010, 11.4144,1.8991, 11.4556,1.9097, 11.4968,1.9149, 
11.5379,1.9165, 11.5791,1.9098, 11.6202,1.8801, 11.6614,1.8127, 11.7026,1.6960, 11.7437,1.5657, 11.7849,1.5000, 11.8261,1.5437, 
11.8672,1.6725, 11.9084,1.8156, 11.9496,1.9077, 11.9907,1.9571, 12.0319,1.9779, 12.0730,1.9954, 12.1142,2.0095, 12.1554,2.0184, 
12.1965,2.0276, 12.2377,2.0355, 12.2789,2.0408, 12.3200,2.0497, 12.3612,2.0556, 12.4024,2.0601, 12.4435,2.0635, 12.4847,2.0666, 
12.5258,2.0707, 12.5670,2.0750, 12.6082,2.0716, 12.6493,2.0774, 12.6905,2.0818, 12.7317,2.0848, 12.7728,2.0873, 12.8140,2.0912, 
12.8552,2.0968, 12.8963,2.1142, 12.9375,2.1257, 12.9786,2.1378, 13.0198,2.1543, 13.0610,2.1677, 13.1021,2.1848, 13.1433,2.1963, 
13.1845,2.2086, 13.2256,2.2141, 13.2668,2.2174, 13.3080,2.2111, 13.3491,2.2115, 13.3903,2.2065, 13.4314,2.2103, 13.4726,2.2158, 
13.5138,2.2161, 13.5549,2.2166, 13.5961,2.2202, 13.6373,2.2217, 13.6784,2.2248, 13.7196,2.2324, 13.7608,2.2410, 13.8019,2.2490, 
13.8431,2.2525, 13.8842,2.2543, 13.9254,2.2548, 13.9666,2.2594, 14.0077,2.2623, 14.0489,2.2655, 14.0901,2.2727, 14.1312,2.2793, 
14.1724,2.2913, 14.2136,2.3044, 14.2547,2.3139, 14.2959,2.3171, 14.3370,2.3215, 14.3782,2.3165, 14.4194,2.3140, 14.4605,2.3113, 
14.5017,2.3109, 14.5429,2.3203, 14.5840,2.3255, 14.6252,2.3319, 14.6664,2.3345, 14.7075,2.3372, 14.7487,2.3429, 14.7898,2.3427, 
14.8310,2.3366, 14.8722,2.3228, 14.9133,2.3063 )}

\thicklines

\put(1,2){\vector(1,0){14}}
\put(1,2){\vector(0,1){6}}

\put(0.7,8.5){$\mu \; [cm^{-1}]$}
\put(15.2,1.8){$x$}

\put(0.8,2.0){\line(1,0){.2}} \put(1.2,2.2){$0.85$}
\put(0.8,6.37){\line(1,0){.2}} \put(1.2,6.57){$1.00$}

\thinlines

\put( 1.00,2){\line(0,-1){0.2}} \put(0.85,1){$0$}
\put( 4.88,2){\line(0,-1){0.2}} \put(4.6,1){$10$}
\put( 8.78,2){\line(0,-1){0.2}} \put(8.5,1){$20$}
\put(12.67,2){\line(0,-1){0.2}} \put(12.39,1){$30\,mm$}

\end{picture}\\
Fig. 8: attenuation profile
\end{center}

\bigskip
\begin{center}
\unitlength0.5cm
\begin{picture}(16,10)
\thicklines


\thinlines

\put(0,1){
\curve(
1.0000,1.8773, 1.0412,1.8982, 1.0823,1.9062, 1.1235,1.9051, 1.1647,1.9153, 1.2058,1.9206, 1.2470,1.9211, 1.2881,1.9237, 
1.3293,1.9225, 1.3705,1.9185, 1.4116,1.9138, 1.4528,1.9156, 1.4940,1.9227, 1.5351,1.9333, 1.5763,1.9458, 1.6175,1.9576, 
1.6586,1.9661, 1.6998,1.9798, 1.7409,1.9886, 1.7821,2.0033, 1.8233,2.0111, 1.8644,2.0119, 1.9056,2.0112, 1.9468,2.0072, 
1.9879,2.0025, 2.0291,2.0046, 2.0703,2.0050, 2.1114,2.0025, 2.1526,2.0055, 2.1937,2.0043, 2.2349,2.0032, 2.2761,1.9995, 
2.3172,1.9978, 2.3584,2.0007, 2.3996,2.0019, 2.4407,2.0016, 2.4819,2.0043, 2.5231,2.0011, 2.5642,2.0058, 2.6054,2.0004, 
2.6465,1.9966, 2.6877,1.9931, 2.7289,1.9792, 2.7700,1.9671, 2.8112,1.9599, 2.8524,1.9601, 2.8935,1.9678, 2.9347,1.9650, 
2.9759,1.9611, 3.0170,1.9623, 3.0582,1.9605, 3.0994,1.9569, 3.1405,1.9654, 3.1817,1.9721, 3.2228,1.9728, 3.2640,1.9786, 
3.3052,1.9814, 3.3463,1.9753, 3.3875,1.9724, 3.4287,1.9643, 3.4698,1.9601, 3.5110,1.9618, 3.5522,1.9582, 3.5933,1.9655, 
3.6345,1.9691, 3.6756,1.9737, 3.7168,1.9799, 3.7580,1.9896, 3.7991,1.9977, 3.8403,2.0096, 3.8815,2.0215, 3.9226,2.0373, 
3.9638,2.0621, 4.0050,2.1055, 4.0461,2.1811, 4.0873,2.3471, 4.1284,2.7151, 4.1696,3.5329, 4.2108,4.8998, 4.2519,6.2081, 
4.2931,6.5000, 4.3343,5.5355, 4.3754,4.1151, 4.4166,3.0492, 4.4578,2.4977, 4.4989,2.2514, 4.5401,2.1405, 4.5812,2.0812, 
4.6224,2.0506, 4.6636,2.0383, 4.7047,2.0288, 4.7459,2.0136, 4.7871,2.0035, 4.8282,1.9973, 4.8694,2.0049, 4.9106,2.0056, 
4.9517,2.0047, 4.9929,2.0028, 5.0340,2.0011, 5.0752,1.9959, 5.1164,1.9902, 5.1575,1.9904, 5.1987,1.9872, 5.2399,1.9869, 
5.2810,1.9802, 5.3222,1.9780, 5.3634,1.9759, 5.4045,1.9658, 5.4457,1.9688, 5.4868,1.9721, 5.5280,1.9785, 5.5692,1.9824, 
5.6103,1.9874, 5.6515,1.9938, 5.6927,1.9912, 5.7338,1.9896, 5.7750,1.9828, 5.8162,1.9773, 5.8573,1.9769, 5.8985,1.9755, 
5.9396,1.9753, 5.9808,1.9769, 6.0220,1.9761, 6.0631,1.9821, 6.1043,1.9904, 6.1455,1.9947, 6.1866,1.9988, 6.2278,2.0022, 
6.2690,2.0005, 6.3101,2.0070, 6.3513,2.0104, 6.3924,2.0129, 6.4336,2.0196, 6.4748,2.0293, 6.5159,2.0346, 6.5571,2.0490, 
6.5983,2.0645, 6.6394,2.0799, 6.6806,2.0975, 6.7218,2.1068, 6.7629,2.1146, 6.8041,2.1185, 6.8452,2.1398, 6.8864,2.1541, 
6.9276,2.1653, 6.9687,2.1800, 7.0099,2.2016, 7.0511,2.2220, 7.0922,2.2455, 7.1334,2.2731, 7.1746,2.2959, 7.2157,2.3214, 
7.2569,2.3539, 7.2981,2.3799, 7.3392,2.4065, 7.3804,2.4384, 7.4215,2.4690, 7.4627,2.5040, 7.5039,2.5416, 7.5450,2.5799, 
7.5862,2.6074, 7.6274,2.6355, 7.6685,2.6601, 7.7097,2.6820, 7.7509,2.6963, 7.7920,2.7245, 7.8332,2.7617, 7.8743,2.8211, 
7.9155,2.8915, 7.9567,2.9863, 7.9978,3.1104, 8.0390,3.2422, 8.0802,3.3051, 8.1213,3.2608, 8.1625,3.1358, 8.2037,3.0032, 
8.2448,2.8976, 8.2860,2.8130, 8.3271,2.7472, 8.3683,2.7035, 8.4095,2.6774, 8.4506,2.6599, 8.4918,2.6435, 8.5330,2.6292, 
8.5741,2.6152, 8.6153,2.5961, 8.6565,2.5745, 8.6976,2.5515, 8.7388,2.5227, 8.7799,2.4965, 8.8211,2.4616, 8.8623,2.4287, 
8.9034,2.3931, 8.9446,2.3585, 8.9858,2.3307, 9.0269,2.3009, 9.0681,2.2748, 9.1093,2.2431, 9.1504,2.2169, 9.1916,2.1982, 
9.2327,2.1764, 9.2739,2.1507, 9.3151,2.1304, 9.3562,2.1189, 9.3974,2.1100, 9.4386,2.1043, 9.4797,2.1027, 9.5209,2.1038, 
9.5621,2.0898, 9.6032,2.0802, 9.6444,2.0696, 9.6855,2.0559, 9.7267,2.0444, 9.7679,2.0334, 9.8090,2.0267, 9.8502,2.0136, 
9.8914,2.0158, 9.9325,2.0052, 9.9737,2.0008, 10.0149,1.9972, 10.0560,1.9899, 10.0972,1.9920, 10.1383,1.9948, 10.1795,1.9935, 
10.2207,1.9894, 10.2618,1.9924, 10.3030,1.9891, 10.3442,1.9847, 10.3853,1.9851, 10.4265,1.9818, 10.4677,1.9828, 10.5088,1.9803, 
10.5500,1.9801, 10.5911,1.9794, 10.6323,1.9773, 10.6735,1.9782, 10.7146,1.9786, 10.7558,1.9830, 10.7970,1.9801, 10.8381,1.9888, 
10.8793,1.9854, 10.9205,1.9799, 10.9616,1.9786, 11.0028,1.9763, 11.0439,1.9751, 11.0851,1.9772, 11.1263,1.9755, 11.1674,1.9804, 
11.2086,1.9801, 11.2498,1.9814, 11.2909,1.9740, 11.3321,1.9704, 11.3733,1.9636, 11.4144,1.9561, 11.4556,1.9614, 11.4968,1.9612, 
11.5379,1.9573, 11.5791,1.9450, 11.6202,1.9092, 11.6614,1.8351, 11.7026,1.7107, 11.7437,1.5724, 11.7849,1.5000, 11.8261,1.5390, 
11.8672,1.6647, 11.9084,1.8050, 11.9496,1.8932, 11.9907,1.9381, 12.0319,1.9537, 12.0730,1.9661, 12.1142,1.9749, 12.1554,1.9784, 
12.1965,1.9823, 12.2377,1.9849, 12.2789,1.9848, 12.3200,1.9883, 12.3612,1.9888, 12.4024,1.9879, 12.4435,1.9859, 12.4847,1.9835, 
12.5258,1.9822, 12.5670,1.9810, 12.6082,1.9721, 12.6493,1.9725, 12.6905,1.9714, 12.7317,1.9690, 12.7728,1.9660, 12.8140,1.9644, 
12.8552,1.9646, 12.8963,1.9769, 12.9375,1.9830, 12.9786,1.9899, 13.0198,2.0012, 13.0610,2.0093, 13.1021,2.0212, 13.1433,2.0275, 
13.1845,2.0345, 13.2256,2.0345, 13.2668,2.0324, 13.3080,2.0205, 13.3491,2.0154, 13.3903,2.0048, 13.4314,2.0031, 13.4726,2.0032, 
13.5138,1.9980, 13.5549,1.9930, 13.5961,1.9912, 13.6373,1.9872, 13.6784,1.9849, 13.7196,1.9871, 13.7608,1.9903, 13.8019,1.9929, 
13.8431,1.9910, 13.8842,1.9874, 13.9254,1.9824, 13.9666,1.9815, 14.0077,1.9789, 14.0489,1.9767, 14.0901,1.9785, 14.1312,1.9797, 
14.1724,1.9865, 14.2136,1.9944, 14.2547,1.9984, 14.2959,1.9962, 14.3370,1.9952, 14.3782,1.9846, 14.4194,1.9766, 14.4605,1.9682, 
14.5017,1.9623, 14.5429,1.9664, 14.5840,1.9663, 14.6252,1.9672, 14.6664,1.9643, 14.7075,1.9616, 14.7487,1.9619, 14.7898,1.9562, 
14.8310,1.9445, 14.8722,1.9249, 14.9133,1.9026 )}

\thicklines

\put(1,2){\vector(1,0){14}}
\put(1,2){\vector(0,1){6}}

\put(0.7,8.5){$\mu \; [cm^{-1}]$}
\put(15.2,1.8){$x$}

\thinlines

\put(0.8,2.98){\line(1,0){.2}} \put(1.2,3.2){$1.2$}
\put(0.8,5.89){\line(1,0){.2}} \put(1.2,6.1){$1.3$}

\put( 1.00,2){\line(0,-1){0.2}} \put(0.85,1){$0$}
\put( 4.88,2){\line(0,-1){0.2}} \put(4.6,1){$10$}
\put( 8.78,2){\line(0,-1){0.2}} \put(8.5,1){$20$}
\put(12.67,2){\line(0,-1){0.2}} \put(12.39,1){$30\,mm$}

\end{picture}\\
Fig. 9: attenuation profile,\\corrected for scattering effects
\end{center}

\onecolumngrid

\newpage
\section{VI. Evaluation and Interpretation}

\noindent
The corrected diffusion profile shows three distinct features:
\\~

\bigskip
\begin{center}
\unitlength0.5cm
\begin{picture}(24,16)
\thicklines


\thinlines

\put(2,3){
\curve(
0.0000,0.7546, 0.0588,0.7963, 0.1176,0.8124, 0.1764,0.8101, 0.2352,0.8306, 0.2940,0.8413, 0.3528,0.8421, 0.4116,0.8475, 
0.4704,0.8451, 0.5292,0.8371, 0.5881,0.8276, 0.6469,0.8311, 0.7057,0.8453, 0.7645,0.8665, 0.8233,0.8915, 0.8821,0.9151, 
0.9409,0.9322, 0.9997,0.9596, 1.0585,0.9771, 1.1173,1.0065, 1.1761,1.0222, 1.2349,1.0238, 1.2937,1.0225, 1.3525,1.0144, 
1.4113,1.0049, 1.4701,1.0093, 1.5289,1.0099, 1.5877,1.0051, 1.6465,1.0110, 1.7054,1.0086, 1.7642,1.0064, 1.8230,0.9989, 
1.8818,0.9955, 1.9406,1.0014, 1.9994,1.0039, 2.0582,1.0033, 2.1170,1.0086, 2.1758,1.0023, 2.2346,1.0116, 2.2934,1.0009, 
2.3522,0.9933, 2.4110,0.9862, 2.4698,0.9585, 2.5286,0.9341, 2.5874,0.9198, 2.6462,0.9201, 2.7050,0.9356, 2.7639,0.9301, 
2.8227,0.9223, 2.8815,0.9246, 2.9403,0.9210, 2.9991,0.9138, 3.0579,0.9309, 3.1167,0.9442, 3.1755,0.9456, 3.2343,0.9572, 
3.2931,0.9629, 3.3519,0.9506, 3.4107,0.9447, 3.4695,0.9286, 3.5283,0.9202, 3.5871,0.9235, 3.6459,0.9165, 3.7047,0.9310, 
3.7635,0.9382, 3.8223,0.9475, 3.8812,0.9598, 3.9400,0.9792, 3.9988,0.9955, 4.0576,1.0191, 4.1164,1.0430, 4.1752,1.0747, 
4.2340,1.1242, 4.2928,1.2111, 4.3516,1.3622, 4.4104,1.6942, 4.4692,2.4302, 4.5280,4.0659, 4.5868,6.7996, 4.6456,9.4162, 
4.7044,10.0000, 4.7632,8.0710, 4.8220,5.2302, 4.8808,3.0983, 4.9396,1.9953, 4.9985,1.5027, 5.0573,1.2810, 5.1161,1.1623, 
5.1749,1.1011, 5.2337,1.0767, 5.2925,1.0576, 5.3513,1.0271, 5.4101,1.0070, 5.4689,0.9947, 5.5277,1.0098, 5.5865,1.0113, 
5.6453,1.0095, 5.7041,1.0056, 5.7629,1.0021, 5.8217,0.9919, 5.8805,0.9804, 5.9393,0.9809, 5.9981,0.9743, 6.0569,0.9738, 
6.1158,0.9603, 6.1746,0.9559, 6.2334,0.9518, 6.2922,0.9316, 6.3510,0.9376, 6.4098,0.9442, 6.4686,0.9571, 6.5274,0.9648, 
6.5862,0.9748, 6.6450,0.9876, 6.7038,0.9825, 6.7626,0.9792, 6.8214,0.9656, 6.8802,0.9546, 6.9390,0.9537, 6.9978,0.9509, 
7.0566,0.9507, 7.1154,0.9539, 7.1742,0.9522, 7.2331,0.9642, 7.2919,0.9809, 7.3507,0.9893, 7.4095,0.9976, 7.4683,1.0043, 
7.5271,1.0011, 7.5859,1.0141, 7.6447,1.0208, 7.7035,1.0257, 7.7623,1.0393, 7.8211,1.0585, 7.8799,1.0691, 7.9387,1.0981, 
7.9975,1.1289, 8.0563,1.1599, 8.1151,1.1950, 8.1739,1.2136, 8.2327,1.2292, 8.2916,1.2369, 8.3504,1.2795, 8.4092,1.3081, 
8.4680,1.3306, 8.5268,1.3601, 8.5856,1.4032, 8.6444,1.4440, 8.7032,1.4910, 8.7620,1.5461, 8.8208,1.5919, 8.8796,1.6427, 
8.9384,1.7079, 8.9972,1.7597, 9.0560,1.8130, 9.1148,1.8767, 9.1736,1.9380, 9.2324,2.0080, 9.2912,2.0832, 9.3500,2.1598, 
9.4089,2.2149, 9.4677,2.2711, 9.5265,2.3202, 9.5853,2.3640, 9.6441,2.3927, 9.7029,2.4490, 9.7617,2.5234, 9.8205,2.6423, 
9.8793,2.7830, 9.9381,2.9726, 9.9969,3.2208, 10.0557,3.4843, 10.1145,3.6102, 10.1733,3.5217, 10.2321,3.2717, 10.2909,3.0064, 
10.3497,2.7952, 10.4085,2.6260, 10.4673,2.4944, 10.5262,2.4070, 10.5850,2.3549, 10.6438,2.3198, 10.7026,2.2871, 10.7614,2.2585, 
10.8202,2.2305, 10.8790,2.1923, 10.9378,2.1490, 10.9966,2.1030, 11.0554,2.0454, 11.1142,1.9931, 11.1730,1.9233, 11.2318,1.8574, 
11.2906,1.7862, 11.3494,1.7170, 11.4082,1.6614, 11.4670,1.6018, 11.5258,1.5496, 11.5846,1.4863, 11.6435,1.4338, 11.7023,1.3964, 
11.7611,1.3528, 11.8199,1.3013, 11.8787,1.2608, 11.9375,1.2378, 11.9963,1.2200, 12.0551,1.2087, 12.1139,1.2053, 12.1727,1.2077, 
12.2315,1.1796, 12.2903,1.1604, 12.3491,1.1392, 12.4079,1.1118, 12.4667,1.0889, 12.5255,1.0667, 12.5843,1.0534, 12.6431,1.0272, 
12.7019,1.0316, 12.7608,1.0103, 12.8196,1.0017, 12.8784,0.9944, 12.9372,0.9799, 12.9960,0.9840, 13.0548,0.9895, 13.1136,0.9870, 
13.1724,0.9788, 13.2312,0.9848, 13.2900,0.9782, 13.3488,0.9694, 13.4076,0.9701, 13.4664,0.9636, 13.5252,0.9656, 13.5840,0.9606, 
13.6428,0.9603, 13.7016,0.9589, 13.7604,0.9547, 13.8193,0.9563, 13.8781,0.9572, 13.9369,0.9661, 13.9957,0.9601, 14.0545,0.9776, 
14.1133,0.9708, 14.1721,0.9598, 14.2309,0.9571, 14.2897,0.9526, 14.3485,0.9501, 14.4073,0.9544, 14.4661,0.9511, 14.5249,0.9609, 
14.5837,0.9601, 14.6425,0.9627, 14.7013,0.9479, 14.7601,0.9407, 14.8189,0.9271, 14.8777,0.9123, 14.9366,0.9229, 14.9954,0.9224, 
15.0542,0.9146, 15.1130,0.8899, 15.1718,0.8185, 15.2306,0.6701, 15.2894,0.4214, 15.3482,0.1448, 15.4070,0.0000, 15.4658,0.0781, 
15.5246,0.3294, 15.5834,0.6100, 15.6422,0.7864, 15.7010,0.8761, 15.7598,0.9074, 15.8186,0.9322, 15.8774,0.9497, 15.9362,0.9568, 
15.9950,0.9646, 16.0539,0.9698, 16.1127,0.9695, 16.1715,0.9766, 16.2303,0.9775, 16.2891,0.9758, 16.3479,0.9717, 16.4067,0.9670, 
16.4655,0.9643, 16.5243,0.9621, 16.5831,0.9441, 16.6419,0.9449, 16.7007,0.9428, 16.7595,0.9380, 16.8183,0.9321, 16.8771,0.9288, 
16.9359,0.9292, 16.9947,0.9538, 17.0535,0.9661, 17.1123,0.9797, 17.1712,1.0024, 17.2300,1.0186, 17.2888,1.0425, 17.3476,1.0550, 
17.4064,1.0689, 17.4652,1.0691, 17.5240,1.0648, 17.5828,1.0409, 17.6416,1.0307, 17.7004,1.0096, 17.7592,1.0063, 17.8180,1.0065, 
17.8768,0.9960, 17.9356,0.9861, 17.9944,0.9824, 18.0532,0.9744, 18.1120,0.9697, 18.1708,0.9742, 18.2297,0.9806, 18.2885,0.9859, 
18.3473,0.9821, 18.4061,0.9748, 18.4649,0.9648, 18.5237,0.9631, 18.5825,0.9578, 18.6413,0.9534, 18.7001,0.9570, 18.7589,0.9595, 
18.8177,0.9729, 18.8765,0.9887, 18.9353,0.9969, 18.9941,0.9923, 19.0529,0.9903, 19.1117,0.9692, 19.1705,0.9531, 19.2293,0.9365, 
19.2881,0.9247, 19.3470,0.9328, 19.4058,0.9325, 19.4646,0.9345, 19.5234,0.9287, 19.5822,0.9232, 19.6410,0.9238, 19.6998,0.9123, 
19.7586,0.8890, 19.8174,0.8499, 19.8762,0.8052 )}

\thicklines

\put(2,2){\vector(1,0){20}}
\put(2,2){\vector(0,1){12}}

\put(1.7,14.5){$\mu \; [cm^{-1}]$}
\put(22.2,1.8){$x$}

\thinlines

\put(.4,12.56){$1.35$}
\put(    2,12.68){\line(-1,0){0.2}} 

\put(.4,9.61){$1.30$}
\put(    2,9.77){\line(-1,0){0.2}} 

\put(.4,6.71){$1.25$}
\put(    2,6.87){\line(-1,0){0.2}} 

\put(.4,3.8){$1.20$}
\put(    2,3.96){\line(-1,0){0.2}} 

\put(    2,3.96){\line(1,0){20}}

\put(    2,2){\line(0,-1){0.2}} \put( 1.85,1){$0$}
\put( 7.56,2){\line(0,-1){0.2}} \put( 7.25,1){$10$}
\put(13.11,2){\line(0,-1){0.2}} \put(12.80,1){$20$}
\put(18.67,2){\line(0,-1){0.2}} \put(17.70,1){$30\,mm$}

\put( 7.2,10.5){a)}
\put(12.2,7.5){b)}
\put(17.2,4.5){c)}

\end{picture}\\
Fig. 10: corrected attenuation profile (enlarged)
\end{center}

\bigskip
\noindent
a)

\bigskip
\noindent
At	$x\approx 28mm$, there is a dip in overall attenuation.
This dip is at the position of the right gap. In fig. 7
this gap appears lighter than the surrounding material.
The drop in attenuation is caused by the gaps between the
adjacent surfaces. As the surfaces are only level to a scale
of about $15 \mu m$ there are tiny hollow gaps of the same size
between the two steel blocks. 
They are filled with air which features an
attenuation which is negligible to that of steel.
The spread of the peak is due
to the limited accuracy to which the sample has been rotated perpendicular
to the beam (projection effect), to the divergence of the neutron beam
(blurring the image) and the finite resolution of the optical system.

\bigskip
\noindent
b)

\bigskip
\noindent
At $x\approx 9mm$, there is a sharp peak in overall attenuation.
The corresponding left gap is visible as a sharp dark 
vertical line in fig. 7. This gap contains a bonding foil
of $50 \mu m$ width. The low level of neutron transmission of
the line is due to the higher boron content in the foil.
Boron is highly visible in neutron radiography because of its
large neutron capture cross-section. The measured attenuation
coefficient is lower than the calculated value, again because
of detection of scattered neutrons. Also, the point spread function
of the detector is wider than $50 \mu m$ thus blurring the peak.

\bigskip
\noindent
c)

\bigskip
\noindent
Finally, the peak at $x\approx 18mm$ is from the TLP bond. The 
attenuation profile shows a peak as wide as that
of the foil in the left slit but less high, accompanied by a much 
wider diffusion-like curve. On the ragiography image in fig. 7 the
middle vertical line indicates the position of the final TLP bond.
The faint grey line is as wide as the left line but less dark. It is
surrounded by a broad halo indicating the diffused boron.

\bigskip
\noindent
The findings lead to the following observations and conclusions:

\bigskip
\noindent
The boron has not been dissolved completely.
Part of it remained at its initial position and
produced the narrow middle peak. It is smaller
however than that of the left slit as most of the
material in fact has diffused into the surrounding metal.

\newpage
~

~

\noindent
The dynamics of this diffusion are governded by Fick's second law $dc/dt = D(T) \cdot d^2c/dx^2$ where $c$ is the concentration
and $D(T)$ is the temperature dependent diffusion coefficient. Two solutions to this equations describe the state
after the first and second stage of the bonding process \cite{fick}:

\bigskip
\noindent
As the foil did not dissolve completely, it resembled a reservoir at $x=0$. In this case the solution to
Fick's law is

\bigskip
$ c \sim e^{ - (x / (2\,\sqrt{D(T)\,t}) )^2}$

\bigskip
\noindent
In fact, the diffusion profile of the center bond, is perfectly fitted by this curve with a characteristic
length scale of $x_0 = 2 \sqrt{D(T)\,t} = 2.4 mm \pm 0.1 mm$:

\bigskip
\begin{center}
\unitlength0.5cm
\begin{picture}(24,16)
\thicklines


\put(1,1.963){
\curve(
1.9623,0.0906, 2.0947,0.1390, 2.2270,0.1195, 2.3593,0.1073, 2.4916,0.0562, 2.6239,0.0149, 2.7562,0.0116, 2.8885,0.0010, 
3.0208,0.0000, 3.1532,0.0121, 3.2855,0.0058, 3.4178,0.0508, 3.5501,0.1136, 3.6824,0.1453, 3.8147,0.1764, 3.9470,0.2018, 
4.0793,0.1895, 4.2116,0.2384, 4.3440,0.2635, 4.4763,0.2822, 4.6086,0.3331, 4.7409,0.4055, 4.8732,0.4455, 5.0055,0.5542, 
5.1378,0.6703, 5.2701,0.7867, 5.4025,0.9188, 5.5348,0.9888, 5.6671,1.0475, 5.7994,1.0764, 5.9317,1.2366, 6.0640,1.3441, 
6.1963,1.4286, 6.3286,1.5394, 6.4610,1.7015, 6.5933,1.8550, 6.7256,2.0317, 6.8579,2.2389, 6.9902,2.4110, 7.1225,2.6021, 
7.2548,2.8471, 7.3871,3.0421, 7.5194,3.2423, 7.6518,3.4820, 7.7841,3.7125, 7.9164,3.9757, 8.0487,4.2583, 8.1810,4.5465, 
8.3133,4.7535, 8.4456,4.9649, 8.5779,5.1497, 8.7103,5.3143, 8.8426,5.4221, 8.9749,5.6340, 9.1072,5.9138, 9.2395,6.3607, 
9.3718,6.8899, 9.5041,7.6026, 9.6364,8.5361, 9.7688,9.5269, 9.9011,10.0000, 10.0334,9.6672, 10.1657,8.7273, 10.2980,7.7297, 
10.4303,6.9357, 10.5626,6.2995, 10.6949,5.8046, 10.8272,5.4761, 10.9596,5.2800, 11.0919,5.1482, 11.2242,5.0250, 11.3565,4.9174, 
11.4888,4.8122, 11.6211,4.6685, 11.7534,4.5059, 11.8857,4.3330, 12.0181,4.1164, 12.1504,3.9196, 12.2827,3.6571, 12.4150,3.4095, 
12.5473,3.1418, 12.6796,2.8815, 12.8119,2.6724, 12.9442,2.4484, 13.0765,2.2520, 13.2089,2.0140, 13.3412,1.8165, 13.4735,1.6758, 
13.6058,1.5122, 13.7381,1.3185, 13.8704,1.1661, 14.0027,1.0795, 14.1350,1.0129, 14.2674,0.9702, 14.3997,0.9575, 14.5320,0.9664, 
14.6643,0.8606, 14.7966,0.7885, 14.9289,0.7091, 15.0612,0.6057, 15.1935,0.5197, 15.3259,0.4364, 15.4582,0.3862, 15.5905,0.2878, 
15.7228,0.3045, 15.8551,0.2243, 15.9874,0.1917, 16.1197,0.1644, 16.2520,0.1098, 16.3843,0.1252, 16.5167,0.1462, 16.6490,0.1366, 
16.7813,0.1057, 16.9136,0.1284, 17.0459,0.1036, 17.1782,0.0704, 17.3105,0.0731, 17.4428,0.0487, 17.5752,0.0562)}

\thinlines

\put(1,1.963){
\curve(
1.9623,0.0419, 2.0947,0.0433, 2.2270,0.0450, 2.3593,0.0471, 2.4916,0.0497, 2.6239,0.0528, 2.7562,0.0567, 2.8885,0.0613, 
3.0208,0.0669, 3.1532,0.0736, 3.2855,0.0816, 3.4178,0.0912, 3.5501,0.1025, 3.6824,0.1159, 3.8147,0.1316, 3.9470,0.1500, 
4.0793,0.1714, 4.2116,0.1963, 4.3440,0.2250, 4.4763,0.2581, 4.6086,0.2959, 4.7409,0.3391, 4.8732,0.3881, 5.0055,0.4435, 
5.1378,0.5059, 5.2701,0.5757, 5.4025,0.6535, 5.5348,0.7399, 5.6671,0.8353, 5.7994,0.9402, 5.9317,1.0550, 6.0640,1.1799, 
6.1963,1.3153, 6.3286,1.4613, 6.4610,1.6178, 6.5933,1.7849, 6.7256,1.9622, 6.8579,2.1494, 6.9902,2.3460, 7.1225,2.5512, 
7.2548,2.7641, 7.3871,2.9839, 7.5194,3.2091, 7.6518,3.4385, 7.7841,3.6706, 7.9164,3.9037, 8.0487,4.1360, 8.1810,4.3657, 
8.3133,4.5908, 8.4456,4.8093, 8.5779,5.0192, 8.7103,5.2185, 8.8426,5.4051, 8.9749,5.5773, 9.1072,5.7331, 9.2395,5.8709, 
9.3718,5.9892, 9.5041,6.0868, 9.6364,6.1624, 9.7688,6.2153, 9.9011,6.2449, 10.0334,6.2508, 10.1657,6.2329, 10.2980,6.1915, 
10.4303,6.1270, 10.5626,6.0402, 10.6949,5.9320, 10.8272,5.8037, 10.9596,5.6566, 11.0919,5.4924, 11.2242,5.3127, 11.3565,5.1194, 
11.4888,4.9146, 11.6211,4.7001, 11.7534,4.4780, 11.8857,4.2503, 12.0181,4.0190, 12.1504,3.7861, 12.2827,3.5533, 12.4150,3.3224, 
12.5473,3.0949, 12.6796,2.8722, 12.8119,2.6558, 12.9442,2.4466, 13.0765,2.2457, 13.2089,2.0538, 13.3412,1.8715, 13.4735,1.6993, 
13.6058,1.5375, 13.7381,1.3863, 13.8704,1.2457, 14.0027,1.1156, 14.1350,0.9958, 14.2674,0.8861, 14.3997,0.7860, 14.5320,0.6952, 
14.6643,0.6132, 14.7966,0.5395, 14.9289,0.4735, 15.0612,0.4147, 15.1935,0.3626, 15.3259,0.3166, 15.4582,0.2762, 15.5905,0.2408, 
15.7228,0.2100, 15.8551,0.1833, 15.9874,0.1602, 16.1197,0.1403, 16.2520,0.1233, 16.3843,0.1088, 16.5167,0.0965, 16.6490,0.0861, 
16.7813,0.0774, 16.9136,0.0701, 17.0459,0.0639, 17.1782,0.0589, 17.3105,0.0546, 17.4428,0.0512, 17.5752,0.0483 )}

\thicklines	

\put(1,2){\vector(1,0){22}}
\put(1,2){\vector(0,1){12}}

\put(0.7,14.5){$\mu-\mu_0 \; [cm^{-1}]$}
\put(23.2,1.8){$x$}

\thinlines

\put(    1,2){\line(-1,0){0.2}} \put(-0.6,1.8){$0.00$}
\put(    1,4.19){\line(-1,0){0.2}} \put(-0.6,3.99){$0.01$}
\put(    1,6.38){\line(-1,0){0.2}} \put(-0.6,6.18){$0.02$}
\put(    1,8.57){\line(-1,0){0.2}} \put(-0.6,8.37){$0.03$}
\put(    1,10.75){\line(-1,0){0.2}} \put(-0.6,10.55){$0.04$}
\put(    1,12.94){\line(-1,0){0.2}} \put(-0.6,12.74){$0.05$}

\put( 1,2){\line(0,-1){0.2}} \put( 0.65,1){$-8$}
\put( 6,2){\line(0,-1){0.2}} \put( 5.65,1){$-4$}
\put(11,2){\line(0,-1){0.2}} \put(10.85,1){$0$}
\put(16,2){\line(0,-1){0.2}} \put(15.85,1){$4$}
\put(21,2){\line(0,-1){0.2}} \put(20.85,1){$8\,mm$}

\put(13.0,12){peak from undissolved bonding interlayer}
\put(12.7,12.2){\line(-3,-2){1.9}}

\put(14.5,9){solution to Fick's law $ \sim e^{-(x/2.4mm)^2} $ }
\put(14.2,9.2){\line(-3,-2){2.25}}

\put(16.0,6){diffusion profile from diffused boron}
\put(15.7,6.2){\line(-3,-2){1.9}}

\end{picture}\\
Fig. 11: attenuation profile and Gaussian fit
\end{center}

\bigskip
\noindent
When comparing the area under the diffusion curve to the
total area unter the attenuation profile of the middle peak, one finds that
about $90\% \pm 5\%$ of the boron has diffused
into the base material.

\bigskip
\noindent
While the interlayer dissolves quickly into the liquid phase, this phase
expands by outward diffusion of the additives leading to further solution
of the solid base material. One could ideally assume that at the beginning
of the second stage of the bonding process there had been a step in
concentration at the liquid-solid interface at $x_{LS}$. In this case the
solution to Fick's law was

\bigskip
$ c \sim 1 - \frac{2}{\sqrt{\pi}} \int \limits_0^{(x-x_{LS})/(2\sqrt{D_{LS}(T) \cdot t})} dy \; \exp(-y^2) $

\bigskip
\noindent
This curve does not fit the data as well as the Gaussian curve did, but the
position of the liquid-solid interlayer can be estimated as $x_{LS} \approx \pm 2.2 mm$
which is on a similar lengthscale as $x_0 \approx x_{LS}$. As expected, diffusion is much
slower here with $D_{LS}(T) \cdot t \approx 0.2 \, ... \, 0.4 \, mm^2$ as compared to
$D(T) \cdot t = 1.44 \, mm^2 \pm 0.12 mm^2$ within the liquid. In any case, the multiple of
diffusion coefficient and time is way too low for a complete diffusion of the interlayer
and subsequent leveling of the gradient in concentration of the additives.

\bigskip
\noindent
From the experimental evidence it appears that the specimen resembles an
imperfect joint created in an incomplete bonding process.
This assumption is supported by the incompletely diffused boron as seen from
the undissolved center peak, as well as from the low value for $D(T) \cdot t$ hinting
to an insufficient diffusion time $t$ or a low diffusion coefficient $D(T)$
caused by a too low temperature $T$.

\newpage
\section{VII. Conclusion and Outlook}

\noindent
Using non-destructive neutron radiography we have visualised a TLP bonding joint
and determined the diffusion profile of the boron contained in the bonding additives.
From this profile we were able to quantitatively determine parameters of the bonding process
and to qualitatively show that the joint had not been made using the
optimum process parameters.

\bigskip
\noindent
Between $5\%$ and $15\%$ of the boron did not resolve, in contrast to the theoretical expectation
of complete diffusion.
The remaining $85\%$ to $95\%$ of the boron
diffused into the surrounding metal with a characteristic diffusion lenght of $2.4mm \pm 0.1mm$
corresponding to a multiple of diffusion coefficient and diffusion time of $D(T) \cdot t = 1.44 mm^2 \pm 0.12 mm^2$.
The diffusion through an idealized liquid-solid interface took place at surfaces $2.2 mm$ apart from the center
with a multiple of $D_{LS}(T) \cdot t \approx 0.3 \pm 0.1$.

\bigskip
\noindent
From these results it is obvious that the sample features an imperfect TLP joint due to a too
low temperature or a too short time at temperature for the interlayer to diffuse completely.
In fact, it was revealed only ex post that the specimen was processed at a temperature below
the recommended brazing temperature. 
This fact and further details of the bonding process below were revealed only after the completion
of the experiment and the evaluation of the data. Thereby we show that
neutron radiography was able to detect the imperfectness of the joint.

\bigskip
\noindent
The sample under investigation was produced as part of a trial to optimise the process parameters of TLP bonding
in terms of brazing temperature and time at temperature. The parameters used for the bonding of the sample under
investigation were the least effective ones compared with higher temperatures at the same time or longer times at the
same temperature. This is supported by complementary methods including tensile and toughness and
SEM micrographs - which fully reflects the non destructive findings that this was not an ideal TLP joint. 

\bigskip
\noindent
This observation highlights the efficacy of neutron radiography for non destructive examination of TLP bonded
joints as it clearly shows that the bond has failed to achieve completion.
These results are also evidence of the fact that neutron radiography begins to move
towards quantitative applications. Already now, neutron radiography and neutron tomography
are proven methods in many aspects of materials science and beyond. We hope to trigger
further interest in this rapidly developing technique and its applications.
Further progress may be expected in a number of areas:

\bigskip
\noindent
Given the rate of data acquisition available on Neutrograph it
could even show the time-resolved dynamics of the bonding process.
The different stages of TLP bonding processes take place on
timescales between seconds and hours. A neutron radiography is produced
from several images taken with an exposure time of some ten milliseconds at
our high flux installation. Therefore it will be possible
to resolve the kinematics of a bonding process both in space and time.
Contrast and hence statistics and time resolution can be enhanced
by substituting the boron in the bonding foil by the highly attenuating
isotope $^{10}$B.

\bigskip
\noindent
Even more information could be extracted by using neutron tomography, in particular for non-flat specimen geometries. As in
medical computer tomography the three dimensional structure of an
object is reconstructed from two dimensional projections. Neutron
tomography has successfully produced reconstructions of objects
featuring a very low contrast \cite{lowcontrast}. This is especially
suitable for late stages of the bonding process where large time
scales enable high counting statistics. On the other hand, high speed neutron
tomography is currently under development and could even resolve the
fast dynamics of the first stages of a TLP bonding process.

\bigskip
\noindent
To summarize, we have non-destructively studied a TLP bonding joint.
We are developing a reliable method which
will provide quantitative, three dimensional, time resolved data of TLP
bonding processes. This data will enable materials science to develop a better
understanding of the bonding process and a further optimization of its parameters
(composition of the foil, temperature curve, ...). We have also shown that this method can be used in 
non-destructive testing to detect incomplete bonding processes.
Finally it also provides difficult to obtain validation data for theoretical modelling of diffusion
processes in metals.

\newpage
\section{VIII. Acknowledgement}

\noindent
The sample was made by The Welding Institute in Abington, UK.
It was provided through the Sheffield Hallam University and
the FaME38 Collaboration at the Institut Laue-Langevin.
The Neutrograph experiment is run in collaboration between
the Institut Laue-Langevin and the University of Heidelberg.
It has been funded by the German Federal Ministry for Research
and Education under contract number 06HD153I.

\end{document}